\documentclass[longauth]{aa} 

\usepackage{longtable,lscape}
\usepackage{amsmath}
\usepackage{amsfonts}
\usepackage{natbib}
\bibpunct{(}{)}{;}{a}{}{,}
\usepackage{graphicx}
\usepackage{multirow}

\usepackage{graphicx}
\usepackage{txfonts}
\usepackage{natbib}
\usepackage[colorlinks=true,citecolor=blue]{hyperref}
\usepackage{color}
\usepackage{xspace}
\usepackage{dcolumn}
\usepackage{longtable}
\usepackage{lscape}
\usepackage{soul}        

\usepackage{rotating}
\usepackage{afterpage}

\newcommand{\MJup}{M$_{\mathrm{Jup}}$\xspace}

\newcommand{\teff}{T$_{e\!f\!f}$\xspace}

\usepackage{txfonts}
\begin{document}

   \title{Narrow belt of debris around the Sco-Cen star HD\,141011 \thanks{Based on observations made with ESO Telescopes at the Paranal Observatory under programs ID 095.C-0607,  097.C-0060, 0101.C-0016, 098.C-0739, and 1101.C-0557.}$^{,}$ \thanks{This work has made use of the the SPHERE Data Centre, jointly operated by OSUG/IPAG (Grenoble),
   PYTHEAS/LAM/CESAM (Marseille), OCA/Lagrange (Nice), Observatoire de Paris/LESIA (Paris),
   and Observatoire de Lyon.}}

   \author{M. Bonnefoy 
          \inst{1}
          \and
          J. Milli\inst{1} 
          \and
          F. Menard\inst{1}
          \and
          P. Delorme\inst{1}
          \and
          A. Chomez\inst{1, 2}
          \and
          M. Bonavita\inst{3, 4}
          \and
          A-M. Lagrange\inst{1, 2}
          \and
          A. Vigan\inst{5}
          \and
          J.C. Augereau\inst{1}
          \and
          J.L. Beuzit\inst{3}
          \and
          B. Biller\inst{3, 4}
          \and
          A. Boccaletti\inst{2}
          \and
          G. Chauvin\inst{1, 6}
          \and
          S. Desidera\inst{7}
          \and
          V. Faramaz\inst{8}
          \and
          R. Galicher\inst{2}
          \and
          R. Gratton\inst{8}
          \and
          S. Hinkley\inst{9}
          \and
          C. Lazzoni\inst{7, 10}
          \and
          E. Matthews\inst{11, 12}
          \and
          D. Mesa\inst{7}
          \and
          C. Mordasini\inst{13}
          \and
          D. Mouillet\inst{1}
          \and
          J. Olofsson\inst{14}
          \and
          C. Pinte\inst{1, 15}
}
  
\institute{Univ. Grenoble Alpes, CNRS, IPAG, F-38000 Grenoble, France
\email{mickael.bonnefoy@obs-univ-grenoble.fr}
\and
LESIA, Observatoire de Paris, CNRS, Université Paris Diderot, Université Pierre et Marie Curie, 5 place Jules Janssen, 92190 Meudon, France 
\and
SUPA, Institute for Astronomy, University of Edinburgh, Blackford Hill, Edinburgh EH9 3HJ, UK 
\and
Centre for Exoplanet Science, University of Edinburgh, Edinburgh, UK 
\and
Aix Marseille Université, CNRS, LAM (Laboratoire d’Astrophysique de Marseille) UMR 7326, 13388, Marseille, France 
\and
Unidad Mixta Internacional Franco-Chilena de Astronomía, CNRS/INSU UMI 3386 and Departamento de Astronom\'{i}a, Universidad de Chile, Casilla 36-D, Santiago, Chile
\and
INAF - Osservatorio Astronomico di Padova, Vicolo della Osservatorio 5, 35122, Padova, Italy 
\and
Steward Observatory, Department of Astronomy, University of Arizona, 933 N. Cherry Ave, Tucson, AZ 85721, USA
\and
University of Exeter, School of Physics \& Astronomy, Stocker Road, Exeter, EX4 4QL, UK
\and
Dipartimento di Fisica a Astronomia "G. Galilei", Universita di Padova, Via Marzolo, 8, 35121 Padova, Italy 
\and
Observatoire de l’Universit\'{e} de Gen\`{e}ve, Chemin Pegasi 51, 1290 Versoix, Switzerland 
\and
Department of Physics, and Kavli Institute for Astrophysics and Space Research, Massachusetts Institute of Technology, Cambridge,
MA02139, USA 
\and
Physikalisches Institut, University of Bern, Gesellschaftsstrasse 6, 3012 Bern, Switzerland
\and
 N\'{u}cleo Milenio Formaci\'{o}n Planetaria - NPF, Universidad de Valpara\'{i}so, Av. Gran Breta\~{n}a 1111, Valpara\'{i}so, Chile
\and
School of Physics and Astronomy, Monash University, Clayton, Vic 3800, Australia 
}

   \date{Received 22/07/2021; accepted 22/10/2021}

 
  \abstract
   {We initiated  a deep-imaging survey of Scorpius-Centaurus A-F stars in 2015. These stars are predicted to host warm inner and cold outer belts of debris reminiscent of the architecture of emblematic systems such as HR\,8799.}
   {We present resolved images of a ring of debris around the F5-type star HD~141011 that was observed as part of our survey. We aim to set constraints on the properties of the disk,  compare them to those of other resolved debris disks in Sco-Cen, and detect companions.}
   {We obtained high-contrast coronagraphic observations of HD~141011 in 2015,  2016, and 2019  with VLT/SPHERE. We removed the stellar halo using angular differential imaging (ADI). We  searched for  scattered light emission from a disk in the residuals and applied a forward-modeling approach to retrieve its morphological and photometric properties. We combined our radial velocity and imaging data to derive detection probabilities for companions co-planar with the disk orientation.}
   {We resolve a narrow ring of debris that extends up to $\sim$1.1" ($\sim$141 au) from the star in the IRDIS and IFS data obtained in 2016 and 2019. The disk is not detected in the 2015 data which are of poorer quality. The disks is best reproduced by models of a noneccentric ring centered on the star with an inclination of $69.1\pm0.9^{\circ}$, a position angle of $-24.6 \pm 1.7^{\circ}$, and a semimajor axis of $127.5\pm3.8$ au. The combination of radial velocity and imaging data excludes brown-dwarf  (M>13.6 \MJup) companions coplanar with the disk from 0.1 to 0.9 au and from 20 au up to 500 au (90\% probability).}
   {HD~141011 adds to the growing list of debris disks that are resolved in Sco-Cen. It is one of the faintest disks that are resolved from the ground and has a radial extent and fractional width ($\sim$12.5\%) reminiscent of Fomalhaut. Its moderate inclination and large semimajor axis make it a good target for the James Webb Space Telescope and should allow a deeper search for putative companions shaping the dust distribution.} 


   \maketitle
%

\section{Introduction}

\begin{table*}[t]
\caption{Log of SPHERE observations}
\label{tab:obs}
\begin{center}
\begin{tabular}{llllllllll}
\hline\hline
			&		&			&	IRDIS &	 IFS & 	& &	&		\\
			\hline
Date, UT start	&	Setup	& Density  &  \multicolumn{2}{c}{$\mathrm{DIT\times NDIT \times N_{EXP}}$}  &  $\Delta$PA & $\epsilon$ &  Airmass	&	$\tau_{0}$ & Remarks	\\
 & 	 &				&	(s)  & (s) & ($^{\circ}$)		&	(")		& & (ms) & 	\\
\hline
25/07/2015, 23:28	& IRDIS-H23+IFS-YJ &	ND\_1.0    & 2$\times$5$\times$1  &   8$\times$3$\times$1 &  0.1 &  1.0 & 1.1 	& 1.4   & Unsat. PSF \\
25/07/2015, 23:33    & IRDIS-H23+IFS-YJ&	ND\_0.0   & 64$\times$3$\times$16  &  64$\times$3$\times$15  & 	 35.7  	&	1.1	& 1.1 & 1.2 & ADI seq. \\	
\hline
09/04/2016, 06:53	& IRDIS-H23+IFS-YJ &	ND\_1.0    & 2$\times$2$\times$1  &   4$\times$1$\times$1 &  0.1 &  0.6 & 1.1 	& 4.6   & Unsat. PSF \\
09/04/2016, 06:54    & IRDIS-H23+IFS-YJ &	ND\_0.0   & 32$\times$8$\times$16  &  64$\times$4$\times$16  & 	 49.2  	&	0.6	& 1.1 & 4.5 & ADI seq. \\	
09/04/2016, 08:05	& IRDIS-H23+IFS-YJ &	ND\_1.0    & 2$\times$2$\times$1  &   4$\times$1$\times$1 &  0.1 &  0.7 & 1.1 	& 3.5   & Unsat. PSF \\
\hline
26/04/2019, 05:22 & IRDIS-BBH+IFS-YJ &	ND\_2.0    & 4$\times$8$\times$1  &   32$\times$1$\times$1  &  0.4 &  0.6 & 1.1 	& 4.3   & Unsat. PSF \\
26/04/2019, 05:25   & IRDIS-BBH+IFS-YJ  &	ND\_0.0   & 32$\times$14$\times$16  &  64$\times$7$\times$16  & 	79.3  	&	 0.6 & 1.1 	& 4.3 & ADI seq. \\	
26/04/2019, 07:31	&  IRDIS-BBH+IFS-YJ&	ND\_2.0    & 4$\times$8$\times$1  &   32$\times$1$\times$1 &  0.3 &   0.6 & 1.1 	& 4.3   & Unsat. PSF \\
\hline
\end{tabular}
\end{center}
\tablefoot{DIT, NDIT, and  $N_{EXP}$ correspond to  the Detector Integration Time per frame, the number of individual frames per exposure, and the number of exposures,  respectively. $\Delta$PA is the amplitude of the parallactic rotation. $\epsilon$ and  $\tau_{0}$ correspond to the seeing and coherence time, respectively.}\\
\end{table*}

The proximity (d=90-200 pc), large number of intermediate-mass stars,  and  young age ($\sim$11-17 Myr) of the Scorpius-Centaurus OB association  \citep[][and references therein]{1999AJ....117..354D} all contribute to make it a niche for the direct-imaging search of young self-luminous planets and circumstellar disks. The planet-finder instruments SPHERE \citep[Spectro-Polarimetric High-contrast Exoplanet REsearch, ][]{2019A&A...631A.155B}  and GPI \citep[Gemini Planet Imager,][]{2008SPIE.7015E..18M} are starting to probe the circumstellar environnements in Sco-Cen at unprecedented contrasts ($10^{-6}$)  down to 0.1" separations (10\,au at 100\,pc). They have already yielded images of 15 debris disks around these stars in the past six years (see Appendix \ref{Appendix:B} and references therein). ALMA resolved eight additional disks in the association \citep{2016ApJ...828...25L, 2017AJ....154..225S, 2017ApJ...849..123M} at more moderate spatial resolutions with no corresponding detection in scattered light so far (HD\,112810, HD\,113766, HD\,121191,  HD\,121617, HD\,131488, HD\,138813, HD\,142446, and HD\,146181). The morphology of  each of these disks  (wing-tilt, rings)  inferred from scattered-light images provides indications about the diversity of planetary system architectures \citep{Lee2016} beyond a separation of 10\,au in a narrow age bin ($\sim$10-17 Myr) immediately after the end of giant planet formation.

We initiated a direct-imaging survey with SPHERE in 2015 to image new giant planets and circumstellar disks around a sample of Sco-Cen F5-A0 stars with high infrared excesses, modeled with two blackbody components, each corresponding to a belt of debris \cite[][hereafter C14]{2014ApJS..211...25C}. This architecture is reminiscent of the iconic systems previously identified by direct imaging, such as  HR~8799 \citep{2008Sci...322.1348M, 2010Natur.468.1080M},  51\,Eridani \citep{2015Sci...350...64M}, the Sco-Cen member HD~95086 \citep{2013ApJ...779L..26R}, and of the Solar System. \cite{Bonnefoy2017} presented the images of two belts of debris  around the F0 star HIP~67497 that we inferred from our survey (Paper I). 

We report in this work the discovery of a narrow ring of debris around the F5V star \object{HD~141011} (\object{HIP~77432}). This intermediate-mass star ($\mathrm{M=1.4\:M_{\odot}}$) is located at a distance of $128.37\pm0.32$ pc according to the Gaia-EDR3 \citep{2020yCat.1350....0G}. The Gaia-EDR3 kinematics and the Banyan $\Sigma$ tool\footnote{http://www.exoplanetes.umontreal.ca/banyan/} \citep{2018ApJ...856...23G} confirm the star's membership in the 16 Myr  old \citep{2002AJ....124.1670M} Upper Centaurus Lupus subgroup \citep{1999AJ....117..354D}.    C14  modeled the infrared excess of the star with two belts of debris: a warm ($499\pm9$ K) belt at 0.8\,au and a cold ($100\pm11$ K)  belt at 25.9\,au. \cite{Ballering2017} proposed their own analysis of the excess, confirming a two-belt architecture with a warm belt at 0.5\,au and a cold belt at a temperature of 101.5\,K with no given location. The authors also reported clear silicate features in the best-fit model spectra, which is indicative of exozodiacal dust. \citet{2015ApJ...808..167J} rather modeled the infrared excess with only one warm belt located at $1.62\pm0.63$ au. These previous modelings of the excess all assumed a distance to HD\,141011 of 96.34\,pc, which has considerably changed with Gaia. They relied mostly on the Spitzer IRS spectrum that was available for that target, which covers a limited fraction of the excess emission and can lead to spurious identification of two-belt components \citep[see section 4.3 of][]{2014MNRAS.444.3164K}.

 Previous adaptive optics observations revealed three candidate companions \citep{2013ApJ...773..170J} with separations between 1.77\arcsec and 4.25\arcsec~, but lacked the sensitivity to resolve the circumstellar emission. Our high-contrast observations allow us to constrain its morphology and to clarify the nature of previously identified candidate companions. We use them as well as new radial velocity observations to search for additional companions. We present the observations and data in \S~\ref{section:data} and an analysis of the disk, candidate companion properties, and detection limits in \S~\ref{section:diskprop}.  We discuss our findings in \S~\ref{sec:discussion}.



\section{Observations}
\label{section:data}
\subsection{SPHERE imaging}
We observed HD~141011  on July 25, 2015,  April 9, 2016, and April 26, 2019, with the VLT/SPHERE instrument \citep{2019A&A...631A.155B} mounted on the VLT/UT3 (Table \ref{tab:obs}).  The instrument was operated  with the IRDIS \citep{2008SPIE.7014E..3LD} and IFS \citep{2008SPIE.7014E..3EC} subinstruments  in parallel. The mode enabled  pupil-stabilized observations of the source placed behind the apodized Lyot coronagraph \texttt{N\_ALC\_YJH\_S} with a radius of 92.5\,mas.  IRDIS  recorded  $\sim11.1\arcsec\times12.4\arcsec$ images  of the target  in the H2 ($\lambda_{c}$=1.593 $\mu$m, width=52 nm) and H3 ($\lambda_{c}$=1.667 $\mu$m, width=54 nm)  filters for the 2015 and 2016 epochs \citep{2010MNRAS.407...71V}. We rather used the broad H-band filter ($\lambda_{c}$=1.625 $\mu$m, $\Delta\lambda$=290 nm) in 2019 in both IRDIS channels to provide a deeper image of the disk. The IFS yielded images with a $1.8\arcsec\times1.8\arcsec$ field of view each in  39 spectral channels covering the 0.96-1.33 $\mu$m spectral range at all epochs. We obtained additional coronagraphic observations of the source with satellite spots  created by a waffle pattern introduced onto the deformable mirror of the instrument for registration purposes. We recorded nonsaturated exposures of the star  placed outside of the coronagraphic mask with neutral density filters (\texttt{ND\_1.0} or \texttt{ND\_2.0}) before and/or after the deep-imaging sequence. The point spread function (PSF) extracted from these images were used to estimate the flux and position of the point sources detected in  the field of view.

\begin{figure*}
\begin{center}
\includegraphics[width=\linewidth]{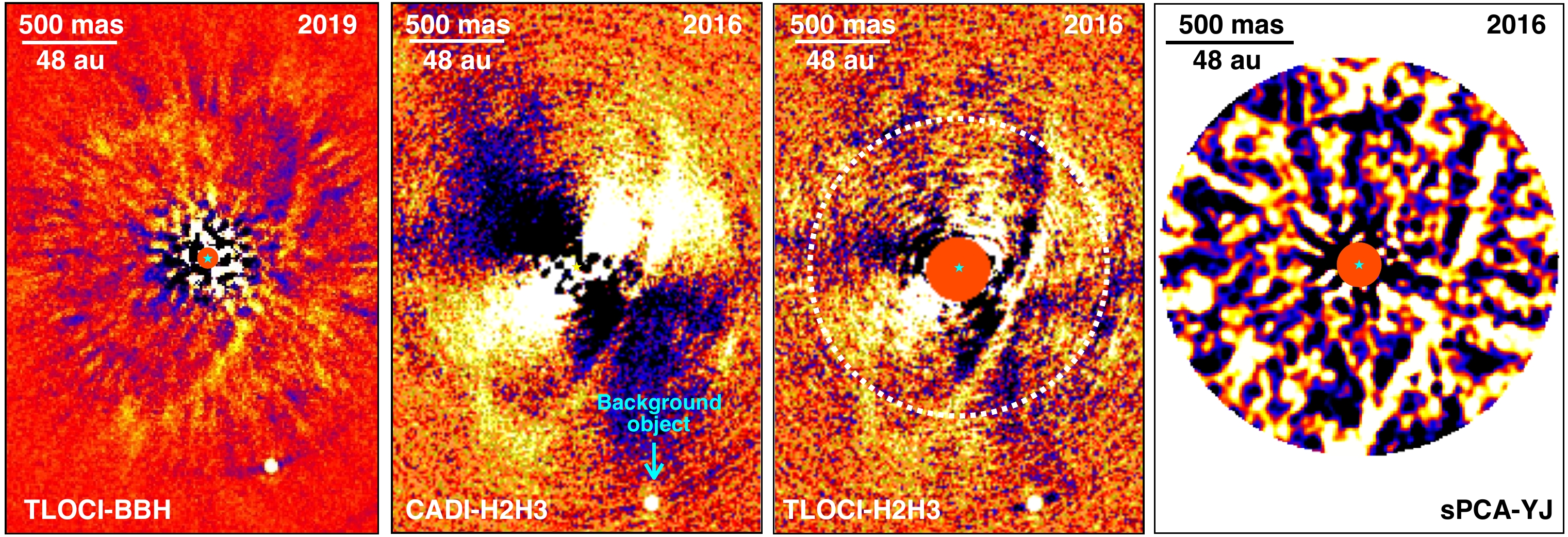}
\caption{Images of the ring of debris around HD\,141011 obtained in 2016 and 2019 with the IRDIS subinstrument and in 2016 with the IFS (convolved with a  Gaussian $\sigma$=2-pixel kernel). The IRDIS images obtained with the cADI and TLOCI algorithms reveal different portions of the ring. The IFS field of view is overlaid (dashed white circle) on the IRDIS/TLOCI panel.}
\label{fig:disk}
\end{center}
\end{figure*}

The data were reduced by the SPHERE Data Center \citep{Delorme2017_DC} in in the same way as for HIP 67497 (Paper I) to obtain a temporal sequence of recentered coronagraphic frames  for each of the 2 IRDIS channels and for the 39 IFS channels. We used true north values of $-1.769\pm0.055^{\circ}$, $-1.731\pm0.070^{\circ}$, $-1.750\pm0.030^{\circ}$ and plate scales of $12.211\pm0.021$, $12.243\pm0.015$, and $12.249\pm0.007$ mas/pixel for the 2015, 2016, and 2019 epochs \citep[see][]{2016A&A...587A..56M}. We adopted a plate scale of $7.46\pm0.02$ mas/pixel for the IFS. 

 We then applied the classical ADI \citep[cADI,][]{2006ApJ...641..556M} and TLOCI \citep{2007ApJ...660..770L} algorithms as implemented in  the \texttt{SPECAL} pipeline of the SPHERE Data Center \citep{Galicher2018_DC} and the  principal component analysis algorithm \citep[PCA;][]{2012ApJ...755L..28S} as implemented in the \texttt{VIP HCI} package \citep{Gomez2017} on the registered sequence of IRDIS frames to supress the stellar halo.  We averaged the H2 and H3 IRDIS images (hereafter H2H3) at each epoch (2015 and 2016)  in order to maximize our sensitivity to faint extended structures. We also averaged the two independent BB\_H frames obtained from the 2019 observations and produced three averaged IFS images from the 39 spectral channels (from Y to J) of the 2015, 2016, and 2019 epochs, respectively. The first five modes of the PCA contained most of the extended structures in the IRDIS images. We also applied the PCA algorithm relying  on the angular and spectral diversity of the IFS data \cite[][hereafter sPCA]{2015A&A...576A.121M}. The  images are shown in Fig. \ref{fig:disk}.  
 We extracted the position and flux of each identified candidate in the TLOCI-reduced frames.  

\subsection{Radial velocity measurements on HIP~141011A}
\label{sec:rvs}
We obtained 13 optical spectra (378-691nm) between March 30, 2018, and March 12, 2020, with the HARPS high-resolution spectrograph \citep{2003Msngr.114...20M} mounted on the ESO 3.6m telescope. The data were reduced using the Software for the Analysis of the Fourier Interspectrum Radial velocities \citep[SAFIR][]{2005A&A...443..337G}. We do not find evidence for spectroscopic binary signals. The measurements are reported in Table \ref{tab:RVs} and are used in Section \ref{section:companions} to set constraints on massive close-in companions in the system.

\begin{table}
\caption{HARPS radial velocity measurements of HD~141011}
\label{tab:RVs}
\centering
\begin{tabular}{c c c}
\hline 
MJD - 2 450 000 & RV (km.s$^{-1}$) & Error (km.s$^{-1}$) \\ 
\hline 
8207.78 &	-0.087 &	0.062 \\
8207.79 &	-0.096 &	0.061\\
8207.81 &	-0.066 &	0.062\\
8208.84 &	-0.021 &	0.038\\
8208.85 &	0.017 &	0.039\\
8534.79 &	0.094 &	0.039\\
8534.80 &	0.061 &	0.040\\
8593.81 &	0.141 &	0.054\\
8593.82 & 0.092 &	0.052\\
8605.71 &	-0.006 & 	0.044\\
8605.72 &	0.113 &	0.043\\
8920.79 &	0.056 &	0.041\\
8920.81 &	0.098 &	0.042\\
\hline
\end{tabular}
\end{table}

\section{Results}
\label{section:diskprop}

We resolve a ring of debris extending up to 1.1" from the star (141.2 au) into the IRDIS field of view in the 2016 and 2019 images (Figure \ref{fig:disk}). The disk is retrieved in the cADI, PCA, and TLOCI reductions. We also detect a similar structure with the IFS data taken in 2016 (Figure \ref{fig:disk} right) and 2019 (marginal detection) when the cubes are collapsed in wavelength. The disk detection is only marginal in the 2015 IRDIS data (no detection in the IFS) because the observing conditions were poorer. We therefore did not use the 2015 images in the remaining paper to characterize the disk.

\subsection{Disk morphology}
\label{sec_morphology}

The various reductions enable us to capture different portions of the disk. The cADI algorithm is the reduction with the least self-subtraction of the disk flux. Therefore it provides deeper images of the faint emission zones far from the star (ansae). The PCA and TLOCI algorithms are more efficient in suppressing the residual halo at short separations. This residual halo is mostly created by the servolag error of the adaptive optics system  \citep[also known as wind-driven halo;][]{Cantalloube2020}.\\ 

To determine the best disk model that best reproduces the scattered-light image of the disk, we used a python implementation\footnote{\texttt{GraTeR} was implemented in python as part of the high-contrast pipeline \texttt{VIP HCI} \citep{Gomez2017}} of the \texttt{GraTeR} code \citep{Augereau1999}. 
The dust radial density is described by a two-component power law of index $\alpha_{in}$ and $\alpha_{out}$, inside and outside a reference radius $r_0$, respectively. The vertical dust distribution is parameterized by a Gaussian profile, with a reference scale height $\xi_0$ at the reference radius $r_0$ and a linear dependence on the distance $r$ to the star (constant opening angle).
The dust density $\rho$ is therefore parameterized with

\begin{equation}
\rho(r,z)=\rho_0 \times \left[ \left( \frac{r}{r_0} \right)^{-2\alpha_{in}} + \left( \frac{r}{r_0} \right)^{-2\alpha_{out}} \right]^{-1/2} \times \exp \left[ -\left( \frac{z}{\xi_0 (r/r_0)} \right)^2 \right]
\end{equation}

We modeled the scattering phase function (hereafter SPF) with a Henyey-Greenstein phase function, which is a common and simple way to describe the phase function of debris disks. It is parametrized by the anisotropic scattering factor $g$ (between $-1$ and $1$),
\begin{equation}
SPF(g,\varphi) =  \frac{1}{4\pi} \frac{1-g^2}{\left( 1-2g\cos{\varphi} +g^2 \right)^{3/2}}
\label{eq_HG}
\end{equation}
where $\varphi$ is the scattering angle. 

\begin{figure*}
\begin{center}
\includegraphics[width=\linewidth]{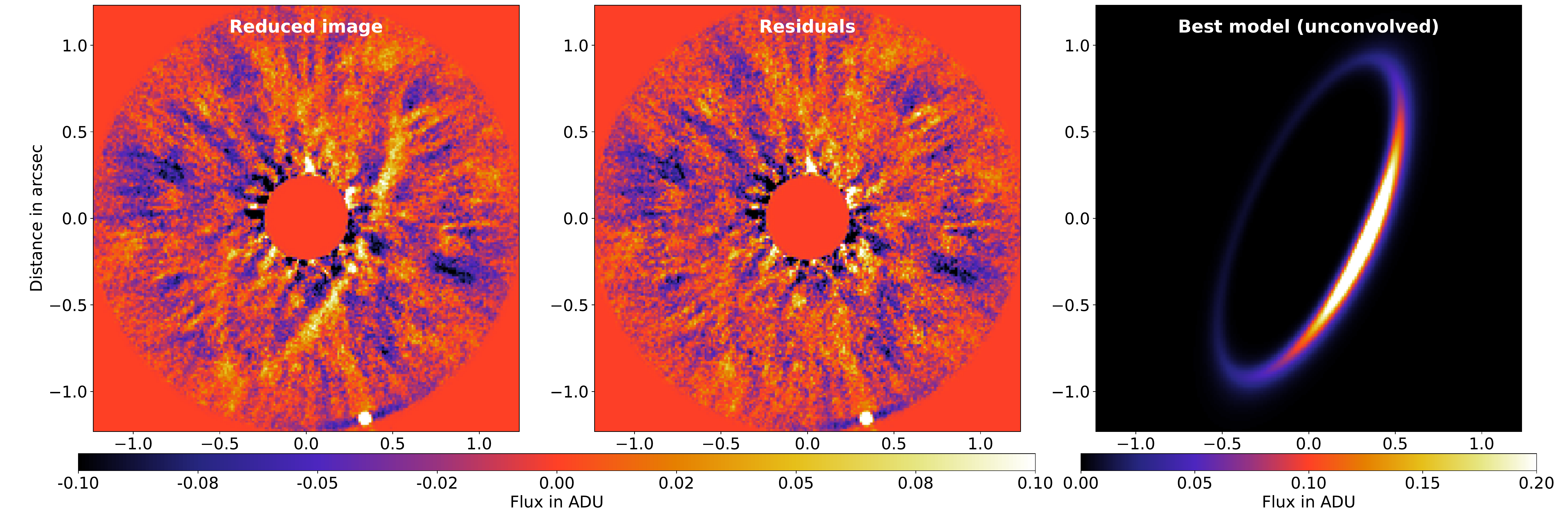}
\caption{Left: IRDIS-reduced image used as a starting point for the forward-modeling, and obtained by averaging the PCA reductions from the 2016 and 2019 epochs. Middle: Residuals after subtraction and forward-modeling of the best model. Right: Best unconvolved model. North is up and east to the left.}
\label{fig_best_model_grater}
\end{center}
\end{figure*}

We applied a forward-modelling approach to find the best disk parameters that are compatible with the data. We first generated a synthetic scattered light image from our model, convolved it with the PSF (here the unsaturated images of the star), and we subtracted it from all individual frames of the IRDIS 2016 and 2019 data sets. We then reduced these disk-subtracted data sets with a PCA algorithm, removing five principal components, to obtain four disk-subtracted images (one for each epoch and IRDIS spectral channel), which we averaged together to have a single disk-subtracted image. We iterated over the disk model parameters until the level of residuals in this final image was minimized. To estimate the goodness of fit, we computed the $\chi^{2}$ in an elliptical ring encompassing the disk. The $\chi^{2}$ was estimated using the following definition of the noise in the final image.  We assumed that the noise is purely radial without azimuthal dependence. The noise radial profile was then estimated as the azimuthal root-mean square of the fluxes encircled in apertures with a diameter of one resolution element ($\lambda/D$), taking into account a correction factor from the small-sample statistics \citep{Mawet2014} and after masking the disk signal in the final image. The uncertainties on the model parameters were derived following the method described in \cite{Bonnefoy2017}.

The initial disk image we started from is shown in Fig. \ref{fig_best_model_grater} (left). 
Initial trials showed that we were little sensitive to the scale height $\xi_0$ as long as it was below 5\,au (for $r_0$ between 125 and 130\,au). We therefore set its value to 3.5\,au to save computational time. This value represents a half-width at half maximum for the vertical density profile of $H=2.9$\,au and an aspect ratio $H/r_0$ of $2.3\%$ that is compatible with the minimum natural aspect ratio of debris disks, which is predicted to be $4\%\pm2\%$ by \citet{Thebault2009}. Similarly, the forward-modeling approach favored models with a very steep inner slope $\alpha_{in}>20$. However, we cannot constrain an inner slope steeper than 20 because we are limited by the angular resolution of our data. We therefore set the inner slope $\alpha_{in}$ to 20. We used as free parameters the reference radius $r_0$, the inclination $i$, the position angle $PA$, the outer slope of the radial dust density $\alpha_{out}$ and the anisotropic scattering factor $g$. We minimized the $\chi^2$ using the \cite{NelderMeade1965} algorithm. 
The best model is shown in Fig. \ref{fig_best_model_grater} (right), along with the residuals after subtraction of the best model (middle). The reduced $\chi^2$ is 1.037, indicating that our model is compatible with the data. The disk peak flux decreases by $\sim 30\%$ due to the convolution with the IRDIS point-spread function, and then by another $\sim 65\%$ due to the ADI reduction technique \citep{Milli2012}. The best parameters of the model are indicated in Table \ref{tab_final_disc_param}. The model favors a narrow ring inclined by $69.1^\circ\pm0.9^\circ$, with a steep outer slope for the radial dust density of $-13.9^{+3.4}_{-4.6}$ and a reference radius $127.5 \pm 3.9$\,au. Because $|\alpha_{in}| > |\alpha_{out}|$, the location of the maximum dust surface density is slightly greater than $r_0$ and equals 128.2\,au, with a ring FWHM of 16.1\,au, corresponding to about three resolution elements in the H band. This implies a fractional width for the ring of 12.6\%.

\begin{table}
\caption{Best model parameters. The uncertainty is given at the $1\sigma$ level.}
\label{tab_final_disc_param}
\centering
\begin{tabular}{c c}
\hline 
\hline 
Parameter & Best value \\
\hline 
$r_0$ (au) & $127.5 \pm 3.9$ \\
$i$ ($^\circ$) & $69.1 \pm 0.9$ \\
$PA$ ($^\circ$) & $-24.6 \pm 1.7 $ \\
$\alpha_{out}$  & $-13.9^{+3.4}_{-4.6}$ \\
$g$ & $0.65 \pm 0.11 $\\
\hline
\end{tabular}
\end{table}

\subsection{Excess emission}
\label{subsec:escess}

\begin{figure}
\begin{center}
\includegraphics[width=\linewidth]{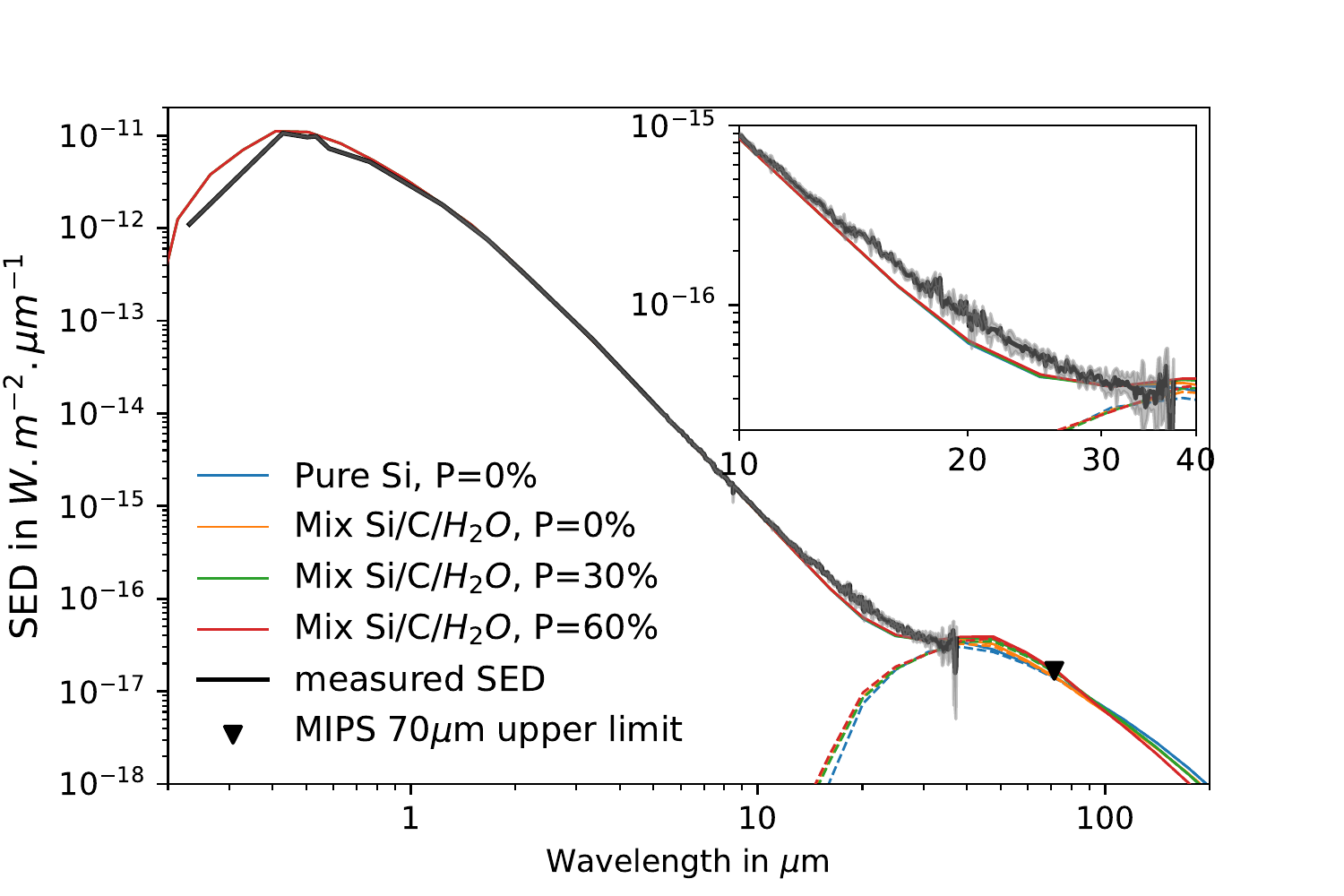}
\caption{Spectral energy distribution of HD\,141011 and the infrared measurements from Spitzer (black line, gray shading for the uncertainty, and triangle for the upper limit). We overplot in colors the four SED models that are compatible with the mid-infrared excess, all with a minimum particle size of $1\:\mu m$. The plain lines represent the stellar photosphere and disk emission, while the dashed lines isolate the contribution of the disk alone. The inset is a zoom in the $10-40\:\mu$m range, highlighting the model mismatch between 15 and $25\:\mu$m.}
\label{fig_sed}
\end{center}
\end{figure}

The parameters of the disk that we detect are not compatible with the previous models of the SED alone \citep[C14; ][]{2015ApJ...808..167J, Ballering2017}. The SED is indeed unconstrained beyond $36\:\mu m$, with only one upper limit from Spitzer/MIPS at $70\:\mu m$ (see Fig.  \ref{fig_sed}). The SED fitting can be highly degenerate in these conditions. The absence of more constraining mid- and far-infrared measurements prevents us from carrying out a detailed modeling of the SED of the cold belt detected in the imaging. We  do not know either whether the infrared excess detected between $10\:\mu m$ and $37\:\mu m$ comes from this cold belt or from a warm belt suggested by C14, \cite {2015ApJ...808..167J}, and \cite{Ballering2017}. 

We therefore modeled the SED anew using the following approach. We assumed that the excess emission detected with Spitzer/IRS between $30\:\mu m$ and $37\:\mu m$ is entirely due to the imaged cold belt. We also investigated which families of dust particle property models are compatible with this hypothesis. 

To do this, we first re-estimated the star's photospheric contribution, normalizing a Kurucz synthetic spectrum with \teff=6500K, log g=4.5, and M/H=0.0 onto a compilation of published photometry (GALEX NUV, TYCHO B, Gaia EDR3, and 2MASS JHK) that is not affected by the excess emission gathered through the VOSA tool \citep{2008A&A...492..277B}. 

We then used the  radiative transfer code MCFOST \citep{Pinte2006, 2009A&A...498..967P} and the disk morphological parameters described in section \ref{sec_morphology} to model the excess emission.

\begin{figure*}
\begin{center}
\begin{tabular}{cccc}
\includegraphics[width=4.3cm]{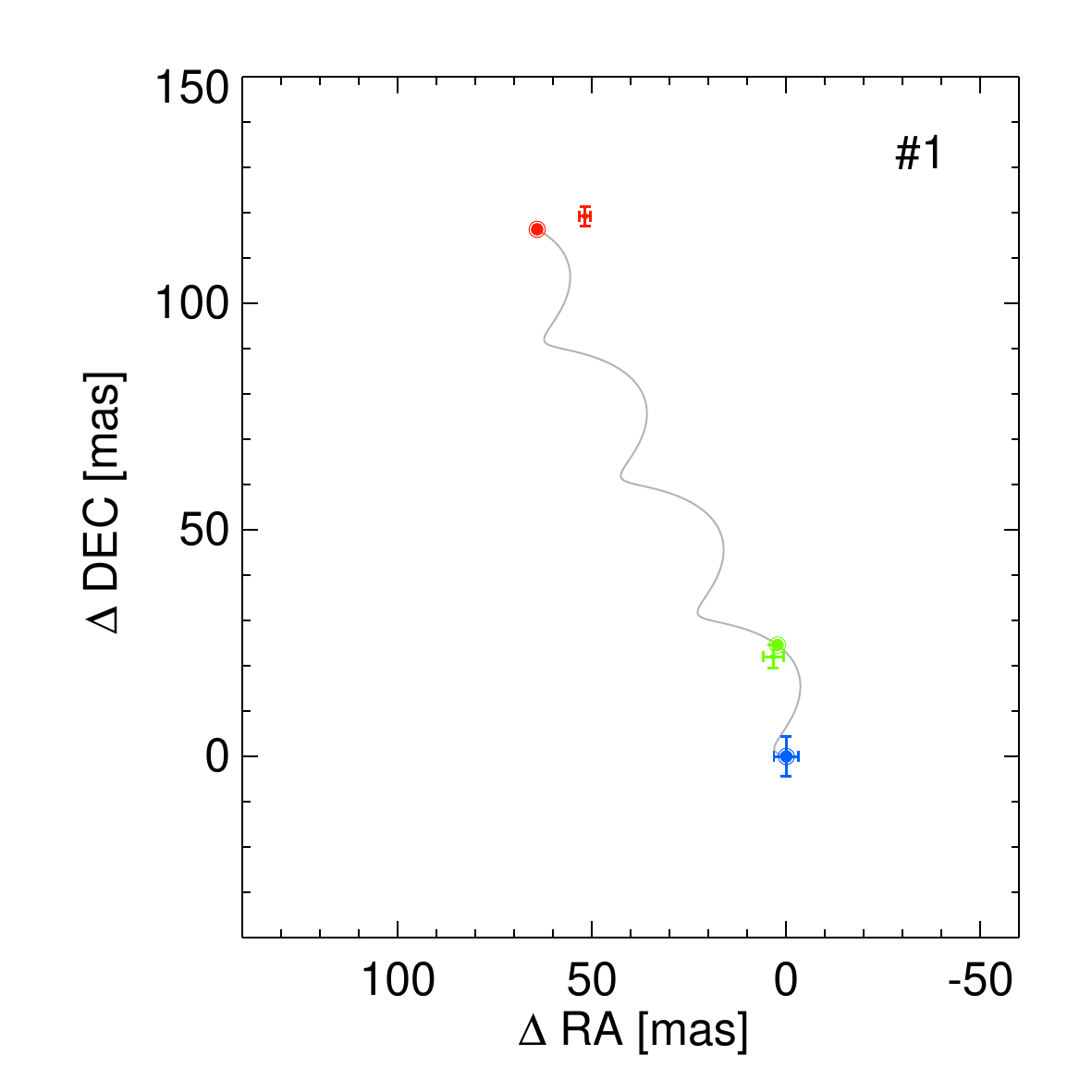}
\includegraphics[width=4.3cm]{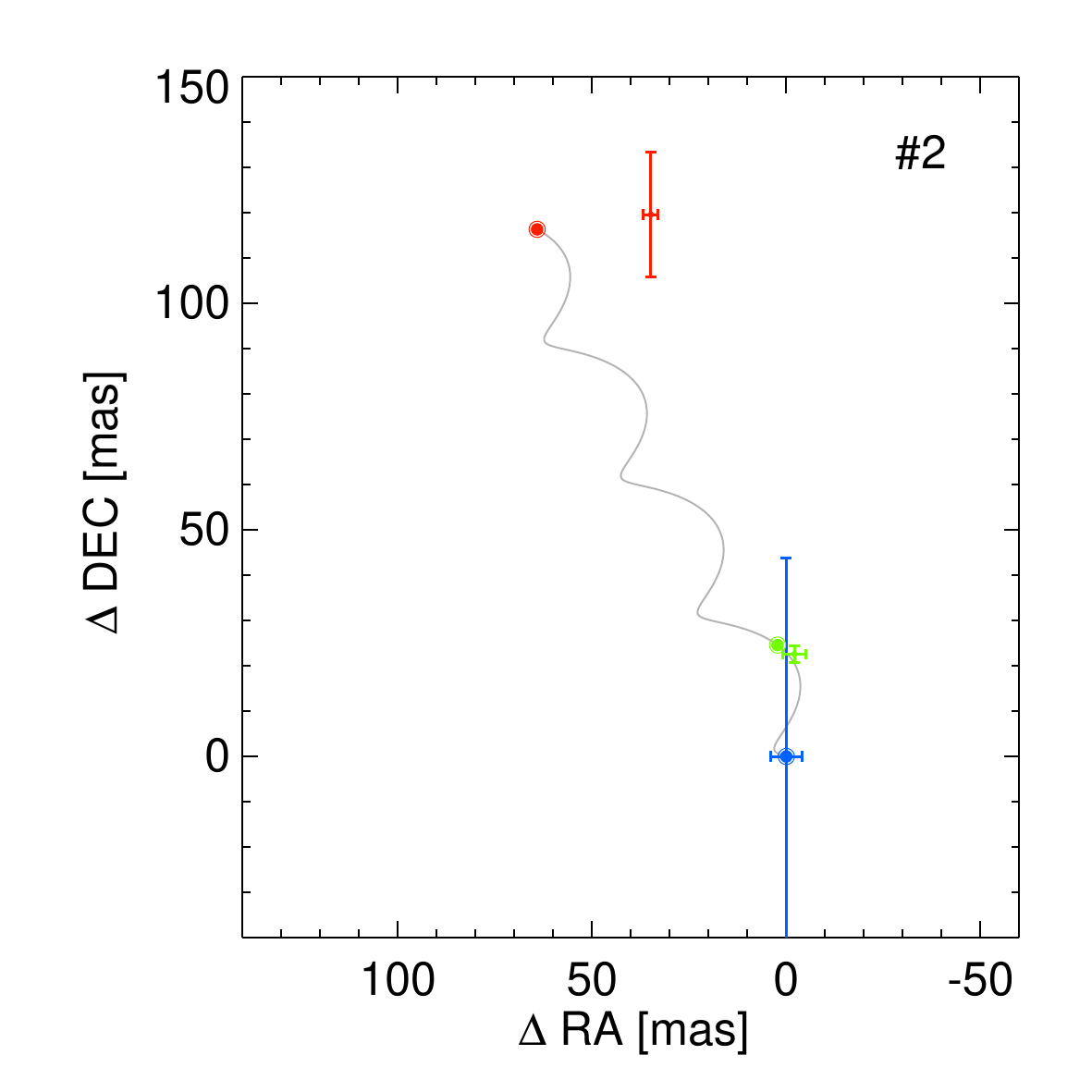}     &  
 \includegraphics[width=4.3cm]{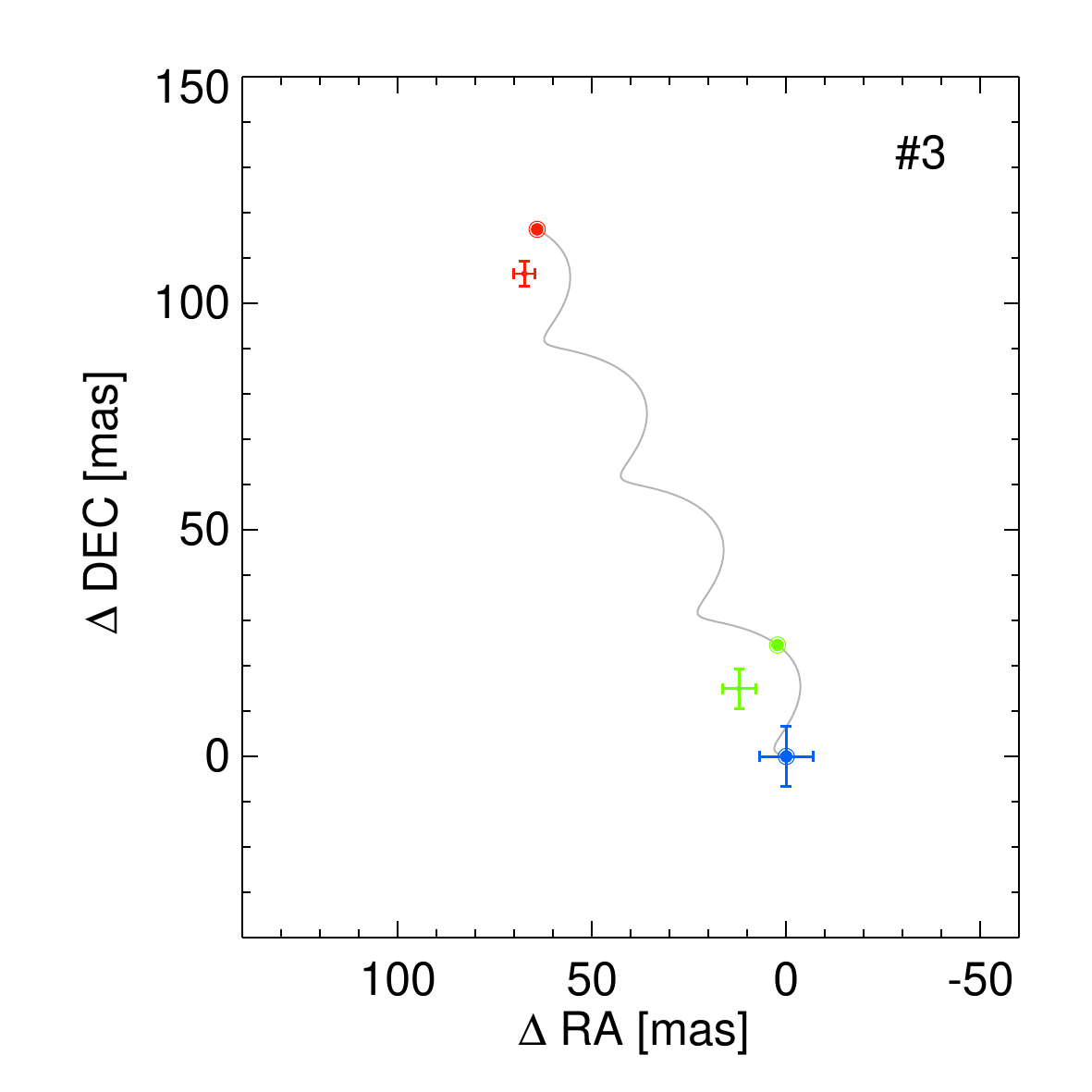}    & 
 \includegraphics[width=4.3cm]{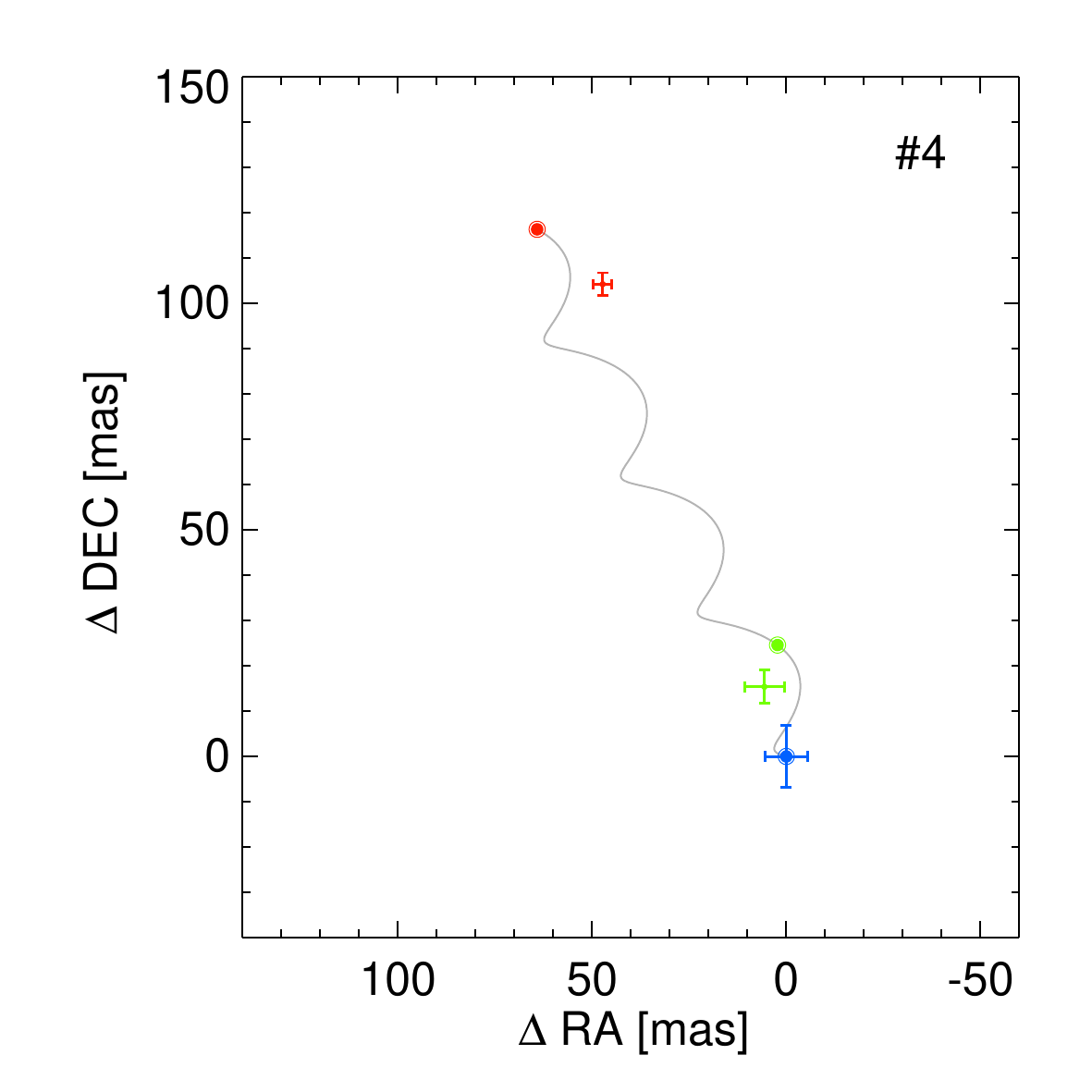}    \\ 
\includegraphics[width=4.3cm]{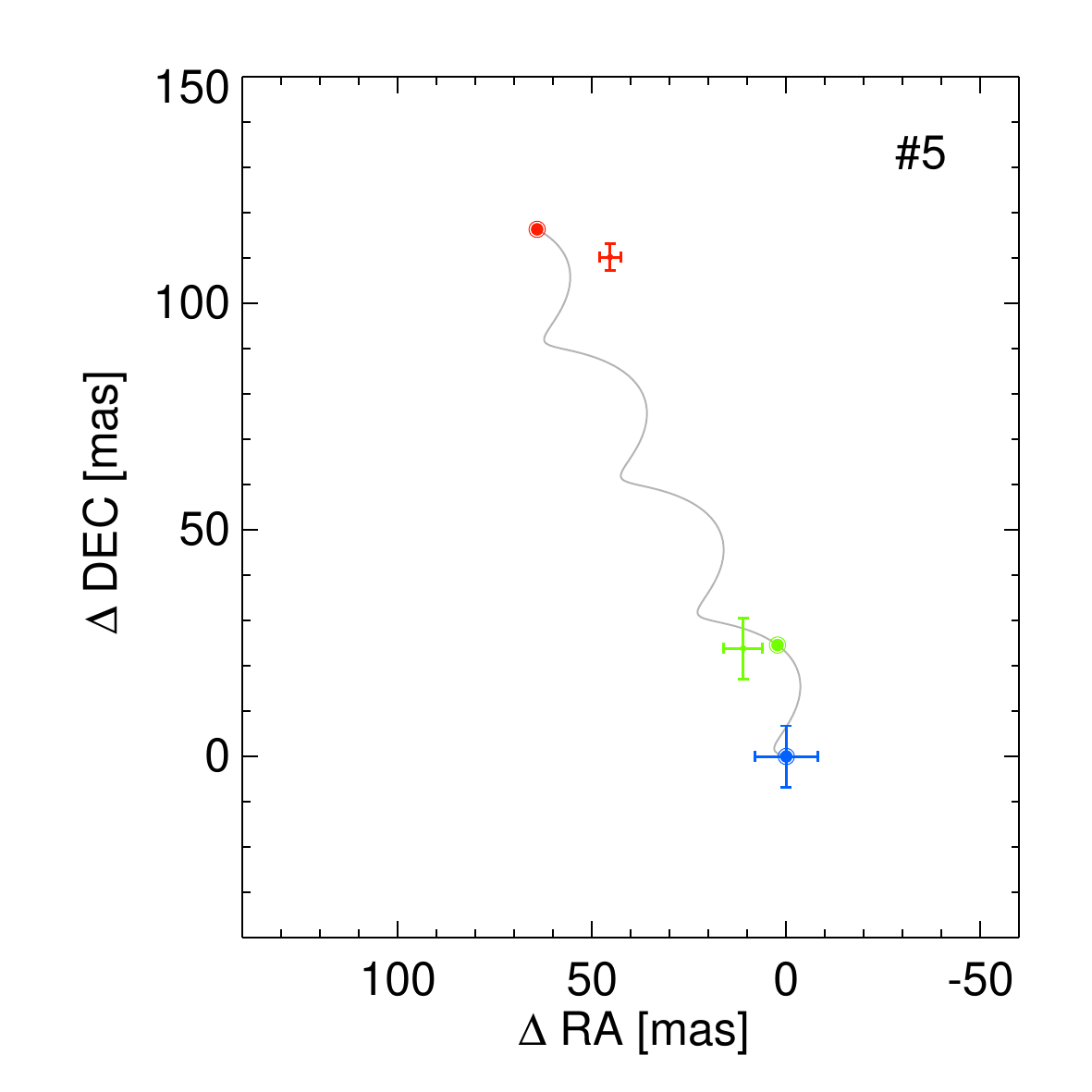}
\includegraphics[width=4.3cm]{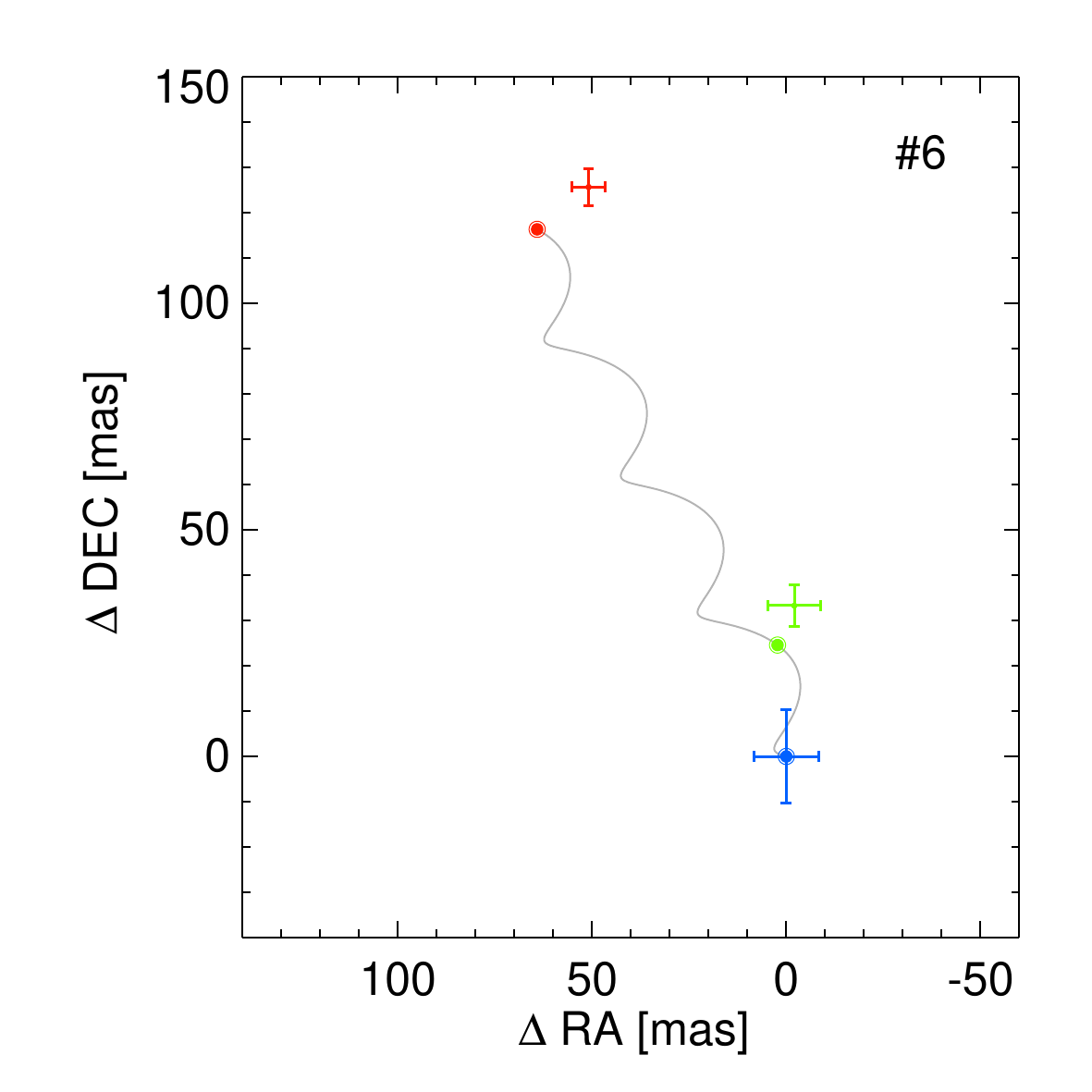}     &  
 \includegraphics[width=4.3cm]{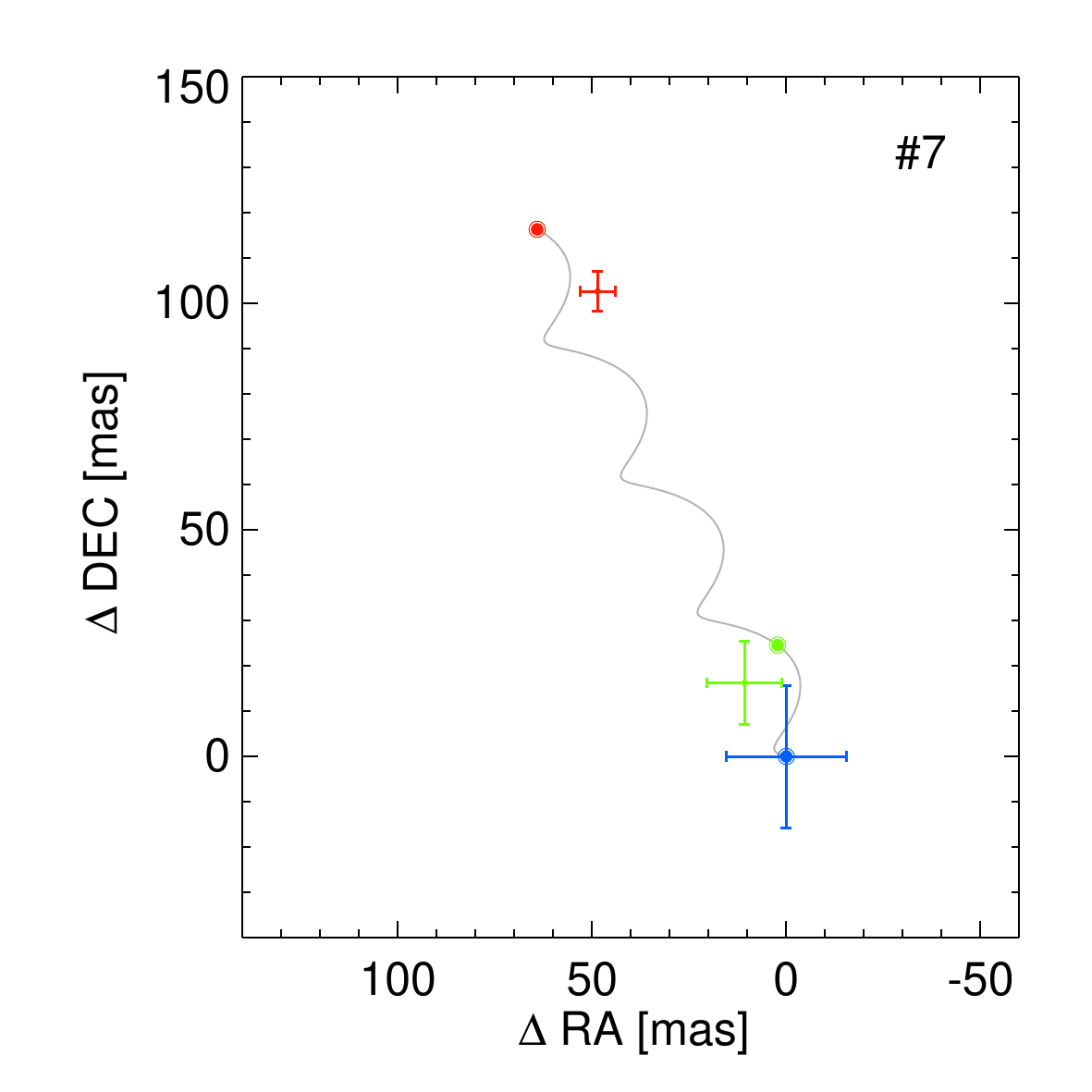}    \\
 
\end{tabular}
\caption{Relative position of the candidate companions in 2015 (blue; epoch 1), 2016 (green; epoch 2), and 2019 (red; epoch 3) with respect to the expected positions (dots) for background objects.}
\label{Fig:cpm}
\end{center}
\end{figure*}

\begin{table}
\caption{Photometry and astrometry of the point sources inferred from the three observation epochs ("Ep."). All are  background stars.}
\label{tab:CC}
\tiny
\begin{center}
\begin{tabular}{llllll}
\hline\hline
\#		&		Ep.$^{a}$			&	$\Delta$H2 or $\Delta$H &	$\Delta$H3  &	 PA & Sep. \\
		&							 & (mag) & (mag)		&	($^{\circ}$) & (mas)	  \\
\hline
1		&		1					&	$12.54\pm0.12$&	$12.33\pm0.12$	&	$196.98\pm0.09$ & $1336\pm4$  \\	
        &       2                   &   $12.48\pm0.07$&	$12.38\pm0.03$	&	$197.12\pm0.09$ & $1314\pm2$  \\	
        &       3                   &   $12.59\pm0.22$ &	\dots	&	$196.28\pm0.04$ & $1207\pm2$  \\	
2		&		1					&	$9.76\pm0.11$ & $9.64\pm0.11$  &	$269.34\pm1.47$ &	$1702\pm3$	 \\
		&		2					&	$9.60\pm0.07$ & $9.50\pm0.07$  &	$270.1\pm0.06$ &	$1704\pm3$	 \\
		&		3					&	$9.62\pm0.22$ & \dots &	$273.43\pm0.47$ &	$1670\pm1$	 \\
3		&		1					&	$11.42\pm0.11$	& $11.32\pm0.11$	&	$132.25\pm0.07$	&	$2900\pm6$  \\	
		&		2					&	$11.34\pm0.07$	& $11.25\pm0.07$	&	$131.87\pm0.08$	&	$2899\pm2$   \\	
		&		3					&	$11.56\pm0.22$	& \dots	&	$129.78\pm0.04$	&	$2881\pm2$  \\	
4		&		1					&	$8.52\pm0.11$	&	$8.43\pm0.11$	&	$157.45\pm0.06$	&	$3358\pm6$ \\				
		&		2					&	$8.33\pm0.07$	&	$8.26\pm0.07$	&	$157.26\pm0.08$	&	$3346\pm2$	 \\				
		&		3					&	$8.58\pm0.22$	&	\dots	&	$155.99\pm0.03$	&	$3281\pm2$		 \\				
5		&		1					&	$9.70\pm0.12$	&$9.68\pm0.12$& 	$68.77\pm0.08$	&	$4329\pm7$		 \\		
		&		2					&	$9.57\pm0.07$	&$9.51\pm0.07$& 	$68.53\pm0.08$	&	$4348\pm3$			 \\		
		&		3					&	$9.73\pm0.23$	& \dots & 	$67.65\pm0.03$	&	$4412\pm2$		 \\		
6		&		1					&	$14.04\pm0.16$	& $13.91\pm0.15$ &	$334.80\pm0.06$	&	$4659\pm9$	\\				
		&		2					&	$13.48\pm0.08$	& $13.50\pm0.91$ &	$334.95\pm0.08$	&	$4690\pm4$	    \\				
		&		3					&	$13.66\pm0.23$	& \dots &	$336.00\pm0.04$	&	$4752\pm3$    \\				
7		&		1					&	$14.42\pm0.28$	& $14.57\pm0.58$ &	$141.52\pm0.12$	&	$4833\pm12$	 \\		
		&		2					&	$14.20\pm0.12$	& $13.92\pm0.09$ &	$141.30\pm0.10$	&	$4827\pm5$		 \\		
		&		3					&	$14.53\pm0.23$	& \dots &	$140.30\pm0.04$	&	$4784\pm3$		 \\		
\hline
\end{tabular}
\end{center}
\tablefoot{$^{a}$ Epoch 1, 2, and 3 correspond to the 2015, 2016, and 2019 observations, respectively.}
\end{table}

We restrained our investigation to a population of dust particles in a size distribution with a power-law index -3.5 \citep{Dohnanyi1969}, with a minimum particle size of 1, 2, 5 and $10\:\mu m$. A rough estimate of the blow-out size $s_{bo}=0.8 \frac{L/L_{\odot}}{M/M_{\odot}} \frac{2700}{\rho}$ \citep[assuming blackbody grains,][]{Wyatt2008} is $1.9\:\mu m$ for $L=3.39\:L_{\odot}$ ($\pm0.03$) and compact astrosilicates of density $\rho=2700\:\text{kg/m}^3$, and more porous particles would have a larger blow-out size \cite[e.g., ][]{Arnold2019}. 

We explored four different compositions: pure astrosilicates \citep{Draine1984}, pure amorphous carbon \citep{Rouleau1991}, a mix of both in equal amounts, or a mix of astronomical silicates, amorphous carbon, and water ice \citep{Li1998} in equal amounts. We used the Mie theory, which is valid for spherical particles and assumed either compact particles or 30\% or 60\% porous particles. We therefore investigated 48 dust models. 

We then started from a low initial mass for the belt and scaled it up until  the SED of the disk and star either matched the Spitzer/IRS measurements between $30\:\mu m$ and $37\:\mu m$, or reached the Spitzer/MIPS $70\:\mu m$ upper limit, whichever occured first. For completeness, we provide in Appendix \ref{app_disk_modelling} the SED of the 48 models we explored. The quality of the fit was assessed using the reduced $\chi^2$ computed between the measured and predicted SED at wavelengths between $30$ and $37\mu m$ (38 degrees of freedom).

Out of our 48 models, 11 are compatible with the $30-37\:\mu m$ IRS spectra and $70\:\mu m$ Spitzer/MIPS upper limit with a reduced $\chi^2$ below 1. These 11 models correspond to dust populations with a minimum particle size of $1$ or $2\:\mu m$  because all larger minimum particle sizes cannot simultaneously be compatible with the infrared excess and the Spitzer/MIPS upper limit. Out of these 11 models, only 4 stay strictly below (but close to) the $70\:\mu m$ Spitzer/MIPS upper limit (the other seven models have a $70\:\mu m$ flux equal to the Spitzer/MIPS upper limit) and these 4 models all have a minimum particle size of $1\:\mu m$ with the best reduced $\chi^2$ below $0.2$.
The MIPS image shows a faint point-source at the target location, but overlaid on a variable background so that it cannot be considered as a detection but may indicate that the disk excess indeed falls close to the tabulated upper limit. We highlight that these four models slightly underpredict the SED between $15$ and $25\:\mu m$, as shown in the inset of Fig. \ref{fig_sed}. We did not try to match the measured flux in this wavelength range because the contribution of the cold belt imaged with SPHERE is negligible compared to a potential warmer belt for which we have no observational constraint from the high-contrast images presented in this paper. A warmer belt at $\sim 500$ K, as proposed by C14, may therefore account for the underestimated $15-25\:\mu m$ flux. We  defer this discussion to a further study  when a far-infrared or (sub-)millimeter SED measurement of the resolved cold belt will be secured to raise the degeneracy between the cold and warmer components. The fractional luminosity of the four best disk models ranges between  $2.7\times10^{-4}$ and $3.0\times10^{-4}$, which is significantly higher than the value reported in C14 ($9.2\times10^{-5}$). 

\begin{figure}
\begin{center}
\includegraphics[width=\linewidth]{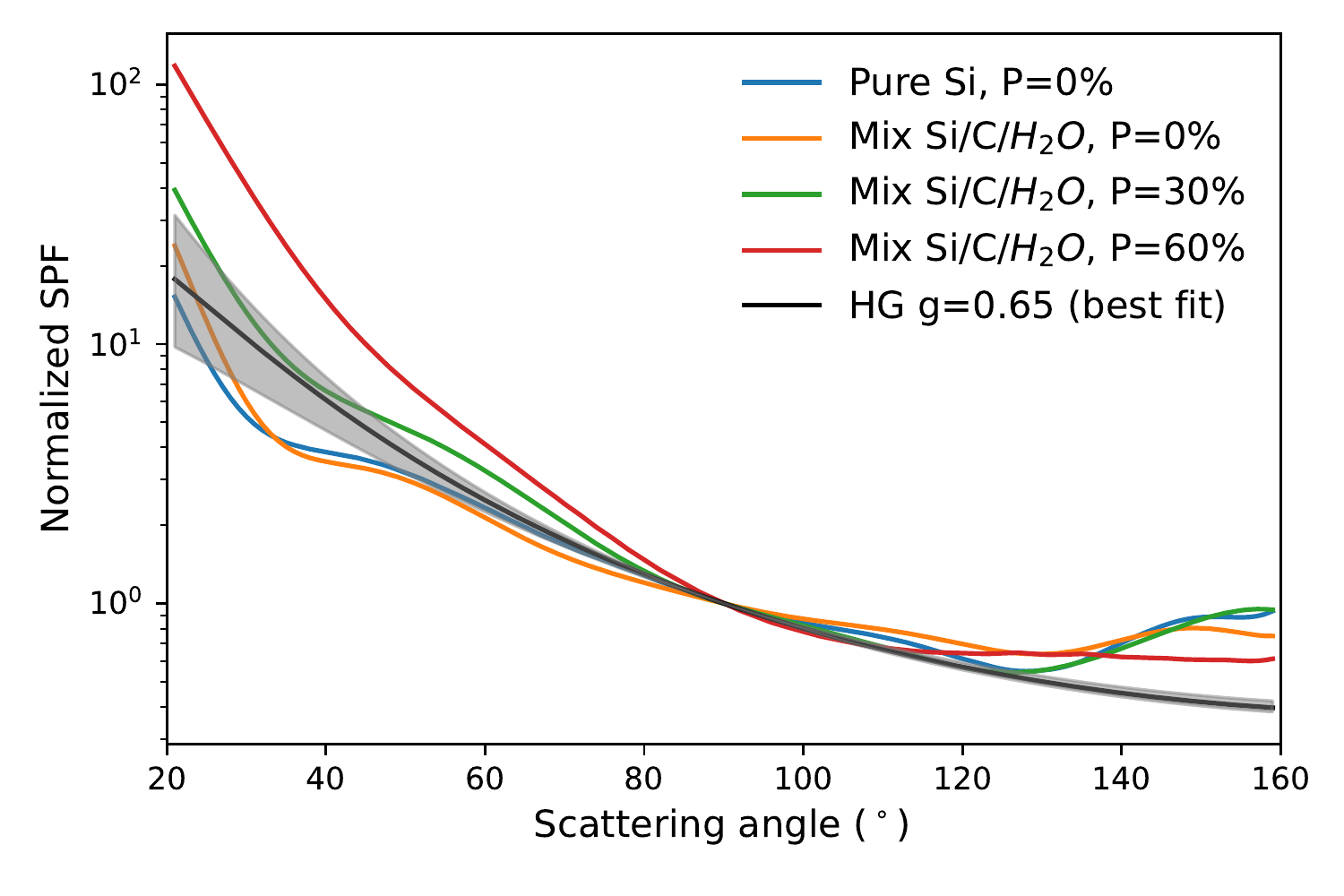}
\caption{Scattering phase functions of the four dust population models that are compatible with the SED (color lines). The black line shows the HG analytical SPF favored by the scattered light forward-modelling, along with the $1\sigma$ uncertainty in gray. All SPFs are normalized to $90^\circ$ for comparison.}
\label{fig:SPF}
\end{center}
\end{figure}

To confirm the compatibility of this family of models with our observations, we verified that they could also reproduce the scattered light images of the disk, especially the scattering phase function (hereafter SPF) of the dust, for which our forward-modeling approach favored an HG with a parameter $g=0.65 \pm 0.11 $. The comparison between the theoretical SPF and the HG model is shown in Fig. \ref{fig:SPF}. It shows that the theoretical SPF are compatible with the fitted SPF, except for the more porous dust population including astrosilicates, carbonaceous material, and water ice. Most models predict some slight backward scattering beyond $140^\circ$. The back side of the disk is unfortunately not detected in the image, and a single HG SPF cannot reproduce this behavior, so that we cannot discuss the reality of this behavior in the present data.  

\subsection{Searching for perturbers}
\label{section:companions}

\begin{figure}
\begin{center}
\includegraphics[width=\linewidth]{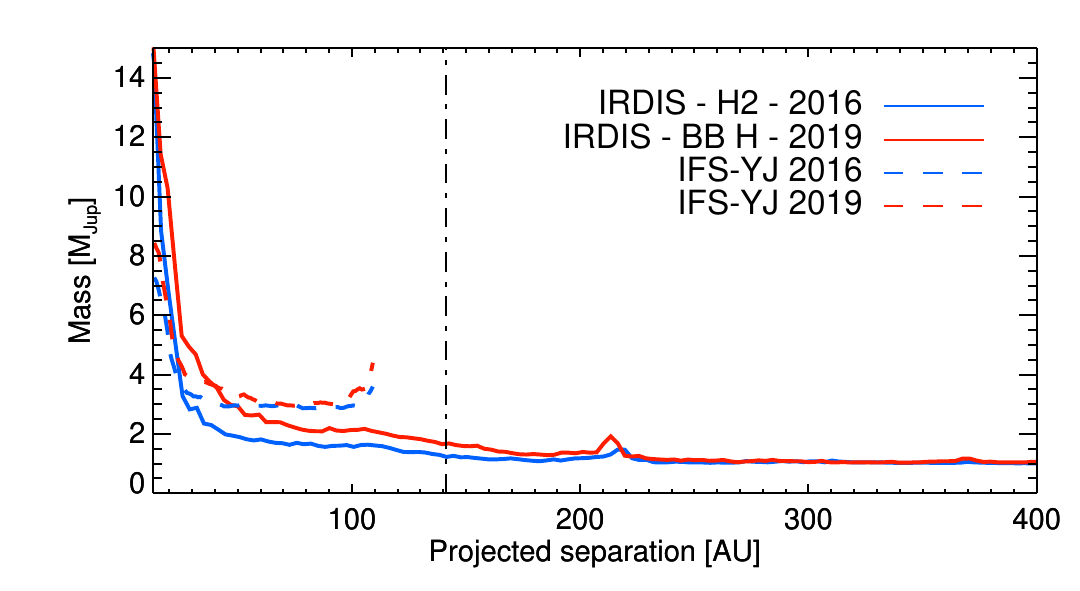}
\caption{Detection limits (5$\sigma$) for the IFS (spectral PCA) and IRDIS (TLOCI) converted into mass assuming the Gaia-EDR3 distance of the system. The vertical dot-dashed line corresponds to the projected separation of the disk ansae.}    
\label{fig:dl}
\end{center}
\end{figure}

We detect seven point sources in  the IRDIS field of view. None of them falls into the field of view of the IFS. Their astrometry is reported in Table \ref{tab:CC} and in Figure \ref{Fig:cpm}. It confirms that all point sources follow the expected apparent on-sky motion of background objects. Point sources 2, 4, and 5 correspond to those detected by \cite{2013ApJ...773..170J}.

Detection limits  for the three epochs of observations were estimated by injecting fake companions with flat spectra into the datacubes for the IFS. The IRDIS detection limits were estimated from the TLOCI coefficients and the local level of the noise.   We used the COND evolutionary tracks to convert the derived contrasts into masses \citep{2003A&A...402..701B} along with the 2MASS photometry of the host star \citep{2003yCat.2246....0C}. We report these limits in Fig. \ref{fig:dl} for the 2016 and 2019 epochs. Any hot-start companion more massive than 2 $M_{Jup}$ would have been detected down to 40 au (projected separation). The IFS data are less sensitive than the IRDIS data beyond 40 au. This is coherent with the loose detection of the disk achieved in the data at J band. The detection limits confirm that the 2019 data are less sensitive than the 2016 epoch in spite of the better seeing and coherence time. 

\begin{figure*}
\begin{center}
\includegraphics[width=\linewidth]{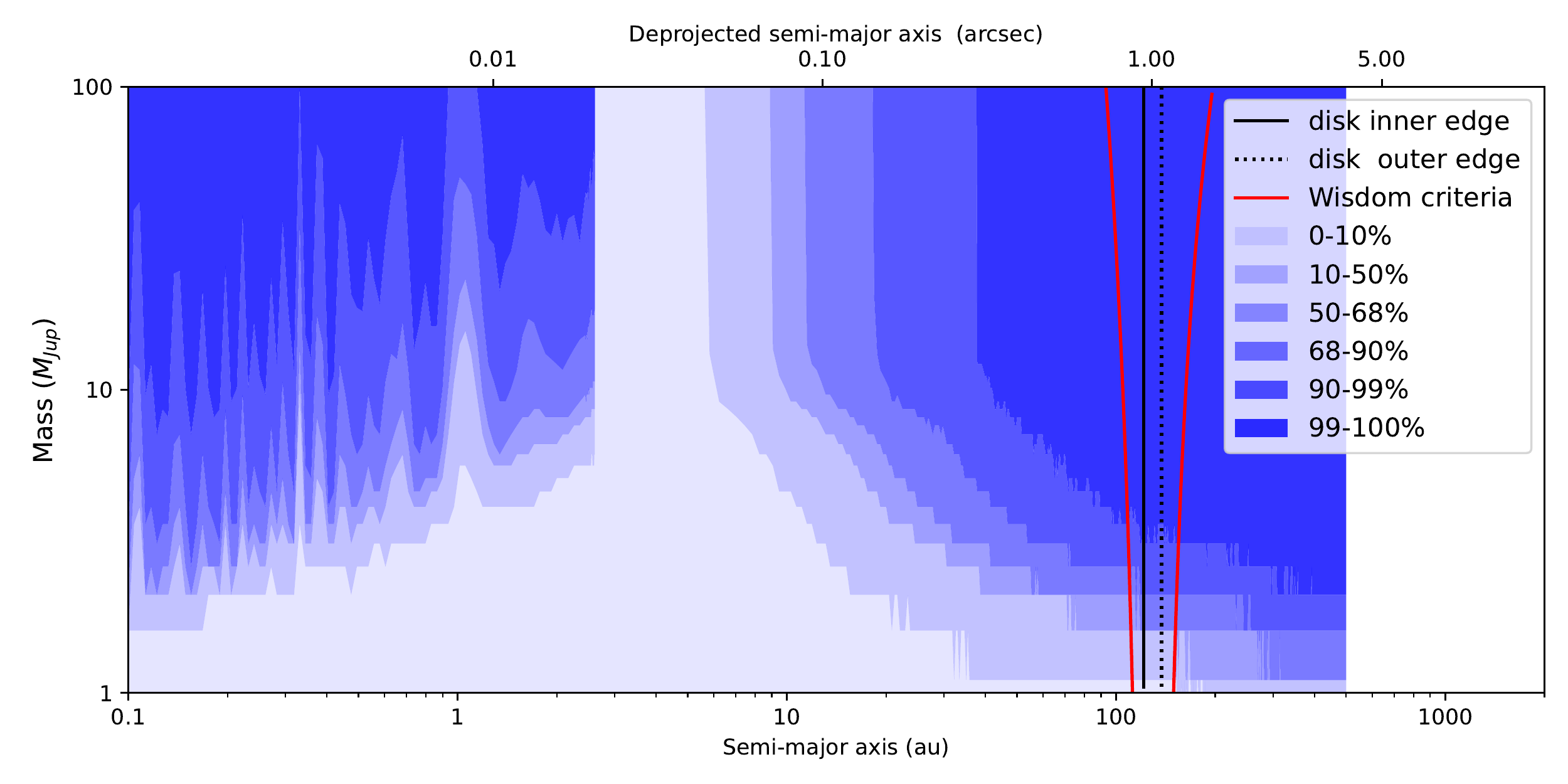}
\caption{Planet detection probabilities inferred from the \texttt{MESS2} tool relying on the available SPHERE and HARPS data of the system. The blue shadings indicate the probability of a detection of an exoplanet of a given mass and semimajor axis. The  vertical black lines indicate the inner and outer edge of the disk. The red line labeled Wisdom criteria indicates the expected mass-to-semimajor axis relation for a planet carving the inner or outer edge of the disk through its chaotic zone.}
\label{fig:alllimdets}
\end{center}
\end{figure*}

We combined the detection limits obtained at the three epochs of SPHERE observations together with the available HARPS radial velocity data (Section \ref{sec:rvs}) on the star through the \texttt{MESS2} tool \citep{2017A&A...603A..54L}. The tool runs Monte Carlo simulations and compares them against the observations at each epoch to evaluate the detection probabilities down to 0.1\,au semimajor axis. We assumed a flat distribution of planet eccentricities from $e=0.0$ to 0.9 and report the results in Fig \ref{fig:alllimdets} for planets coplanar with the disk. 

We exclude (90\% probability) brown dwarf (M$>$13.6~\MJup) companions from 0.1 to 0.9 au, for instance, inside the warm belt component proposed by C14. The imaging data rule out planets more massive than 4 \MJup~beyond 50 au and up to the disk inner edge. 

In an attempt to bring additional constraints from the known system architecture, we investigated whether the very steep inner and outer edges of the disk might be due to the presence of an undetected low-mass companion.
A possible mechanism to explain such a sharp ring is indeed the presence of a gravitational perturber shaping the inner or the outer edge of the disk because a planet is surrounded by a chaotic zone where the orbits of potential particles are unstable \citep{Wisdom1980}. The width $\delta a$ of this chaotic zone created by the overlap of mean-motion resonances of a planet on a circular orbit is given by 
\begin{equation}
\delta a/a = 1.3\mu^{2/7} 
\label{eq:chaotic_zone}
\end{equation}
where $\mu$ is the ratio of the planetary to stellar mass, and a is the semimajor axis of the planet. \citet{Mustill2012} extended this result with numerical simulations to the case of a planet on an eccentric orbit, showing that Eq. \ref{eq:chaotic_zone} still applies when the eccentricity is kept below the critical value of $0.21\mu^{3/7}$. In our case, this means that Eq. \ref{eq:chaotic_zone} remains valid for an eccentricity below $0.04$ for planets with masses lower than 30 \MJup. We show in Fig. \ref{fig:alllimdets} with the red line the mass and semimajor axis of planets with an outer (inner) edge of the chaotic zone corresponding to the inner (outer) edge of the debris disks, overplotted with our detection limits. We can exclude with a confidence level higher than 90\% that planets more massive than 3 \MJup~shape the inner or outer edge of the disk.  

To conclude, we note that the analysis of the Gaia-eDR3 and HIPPARCOS data  \cite[e.g, ][]{2021arXiv210910912K} do not reveal significant differences in proper motion values that could be exploited to place further constraints on unseen companions in the system. 

\section{Discussion and conclusions}
\label{sec:discussion}
\begin{figure}
\begin{center}
\includegraphics[width=\linewidth]{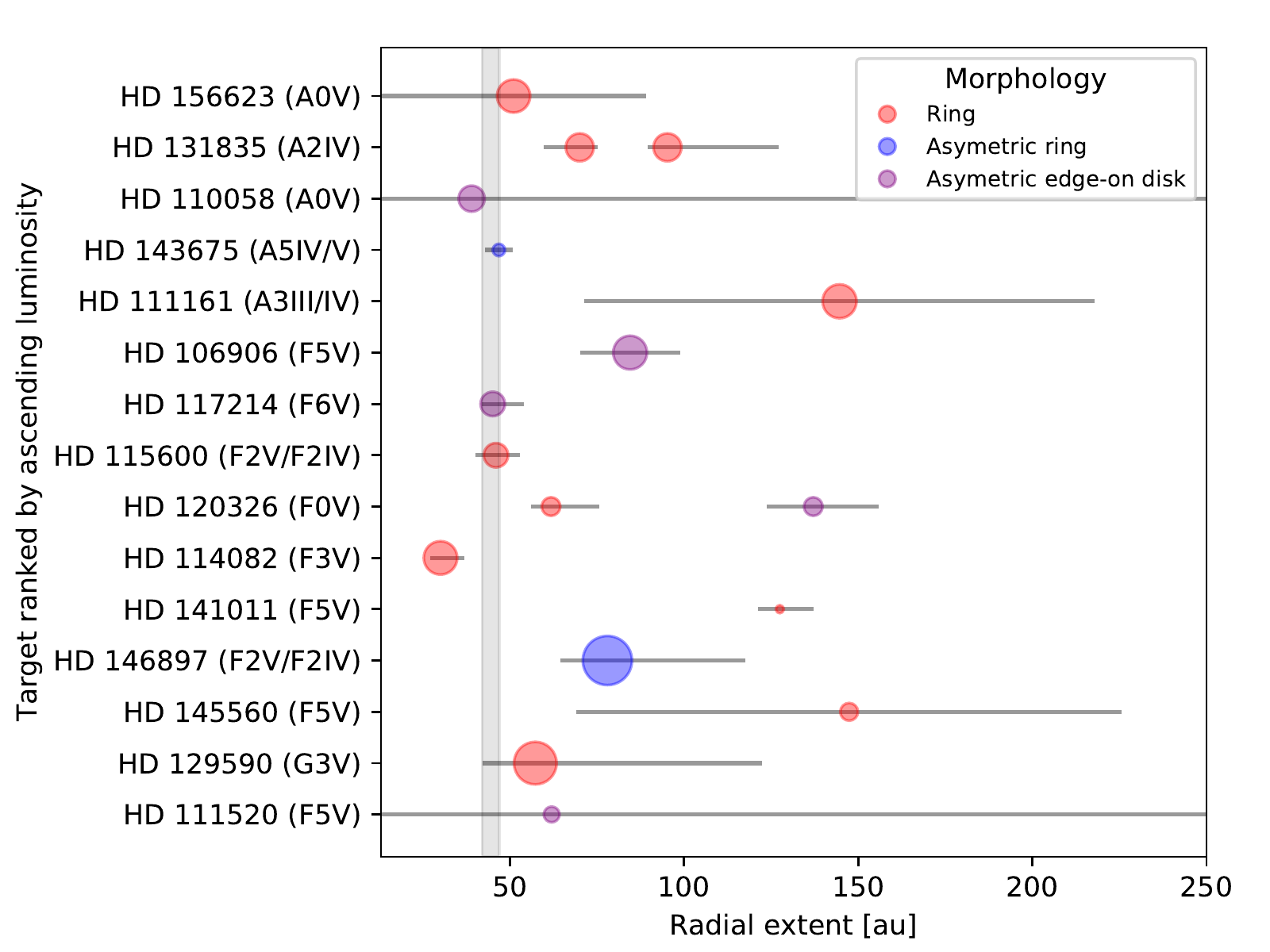}
\caption{Radial extent of the debris disks in Sco-Cen inferred from scattered-light images obtained with SPHERE and GPI. The systems are ranked by stellar luminosities. The symbols are located at the peak of the brightness distribution ($\mathrm{r_{peak}}$) or at the location of estimated $\mathrm{r_{0}}$, and their sizes are proportional to the fractional luminosity. The figure does not extend below 13 au, which corresponds to the typical shortest physical separations accessible by SPHERE and GPI for Sco-Cen targets. The  extent of the cold Kuiper belt is reported for comparison (shaded area).}
\label{fig:morph_lum}
\end{center}
\end{figure}

HD\,141011 is the 53th debris disk that has been resolved in scattered light. It can be compared to the growing sample of related GPI and SPHERE disks that are resolved around stars from the association (see the Appendix \ref{Appendix:B} for a summary of the disk and host star properties based on the latest observational inputs, e.g., the Gaia eDR3). We show in figure \ref{fig:morph_lum} the radial extent of these disks (except for HD98363, for which no values are reported in the literature). We overlaid the position of the cold classical Kuiper belt (42-47 au) whose objects are thought to have experienced the least interactions with the migrating planets \citep{2010ApJ...722L.204P, 2021Icar..35714121G} and might provide a representative comparison to the radial dust distribution of young debris disks.  The symbol sizes are proportional to the reported fractional luminosities.  

\begin{figure}
\begin{center}
\includegraphics[width=\columnwidth]{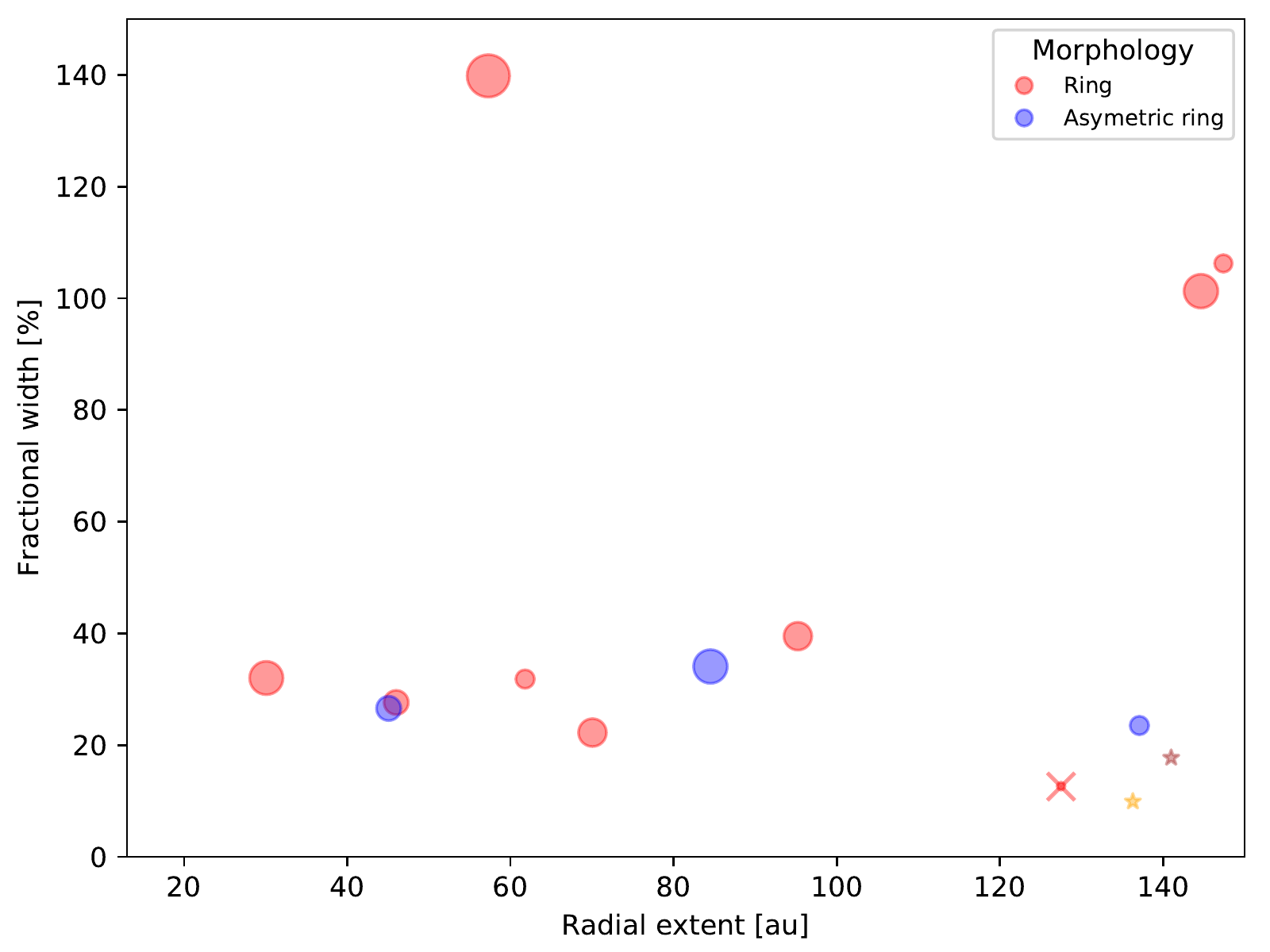}
\caption{Fractional width and radius of debris rings with non-edge-on geometries in Sco-Cen (circles). The circle sizes are proportional to the disk fraction luminosity. HD~141011 (cross) falls in a void space close to the outer belt of HD~120326. The morphology of Fomalhaut inferred from optical  \citep[brown star,][]{2005Natur.435.1067K} or ALMA \citep[yellow star,][]{2017ApJ...842....8M} data are similar to those of HD~141011.}
\label{fig:fracR_R}
\end{center}
\end{figure}

HD141011 has the faintest fractional luminosity ($L_{IR}/L_{\star}$) and is one of the largest rings of debris that have been imaged in Sco-Cen.  Its inclination and brightness enhancement on the forward-scattering side makes it favorable for detection in scattered light, and our multi-epoch observations have been critical in vetting the detection. The figure also shows no clear correlation between the belt spatial location and the star's luminosity \citep[e.g., ][]{2014ApJ...792...65P, 2018ApJ...859...72M, Esposito2020}. The sample in Sco-Cen is still  limited (in particular, the range of stellar luminosity), however, and the figure rather reveals the unique diversity of architectures encountered in the association. 

The ring is found at a larger separation than predicted by the power-law radius - stellar luminosity relation  derived by \cite{Esposito2020} and \cite{2018ApJ...859...72M}. When  resolved debris disks are considered regardless of their age, the reference radius $r_{0}$ and fractional luminosity of HD~141011 are relatively close to those of the ring of debris surrounding  the F-type star HD\,160305 \citep{2019A&A...626A..95P}. This  young \citep[$\sim18-26$ Myr; e.g., ][]{2020A&A...642A.179M} disk has also been resolved with SPHERE, but shows a clear asymmetry, and the outer edge is shallower than that of HD~141011.

The fractional width of the rings ($\Delta R/R$) can inform about the dust confinement mechanisms \citep{2012MNRAS.419.3074M}. The disks in Sco-Cen have all been observed at commensurable spatial resolutions (SPHERE/GPI). However, large uncertainties remain on the disk $R_{in}$ and $R_{out}$ values (Table \ref{tab:dscocen}) because most of these values rely on the modeling of the flux distribution (with sometime fixed $\alpha_{min}$ values) in a relatively low S/N regime, which in turn depends on the disk orientation. A tentative analysis is shown in Figure \ref{fig:fracR_R}. Disks with radii smaller than 100 au have comparable fractional widths below $\sim$40\%. HD129590 appears as an outlier in this separation range: it has the largest fractional width and highest fractional luminosity. \cite{2017ApJ...843L..12M}  noted however that the morphology of this disk  might be more complex than a simple ring. Moreover, we did not include several disks whose morphology is less constrained (HD 156623, HD146897) and which might populate this parameter space.  Conversely, HD141011 falls close to the outer belt of HD 120326, but the latter is a peculiar case, showing a strong asymmetry that might be related to large-scale structures seen at optical wavelengths \citep[]{Bonnefoy2017}. When  all known debris disks resolved in scattered light are considered  \citep[see][for a complete sensus]{2021arXiv210706316A}, the spatial extent and fractional width of HD~141011 are in fact  similar to the ones of the emblematic older disk around Fomalhaut \citep{2005Natur.435.1067K, 2013ApJ...775...56K, 2017ApJ...842....8M}. The fractional width and low (or null) eccentricity of HD141011 is also similar to that of HR4796, whose age is more similar to that of HD~141011 \citep[e.g., TW Hydrae association; $\sim$10 Myr][]{2013ApJ...762..118W, 2014A&A...563A.121D}. Planets have been proposed to explain the morphologies of both Fomalhaut and HR4796 \citep[e.g.,][]{2012A&A...546A..38L, 2021arXiv210304977P}, and the detection limits (Section \ref{section:companions}) for the system must now be improved, in particular, along the semiminor axis with deeper observations benefiting from improved AO and contrast performances \citep[SPHERE star-hopping mode, SPHERE+, see][]{2020arXiv200305714B}, the JWST, or the E-ELT.  


\begin{acknowledgements}
We thank our anonymous referees for their constructive comments which helped improving the manuscript. We thank the ESO staff for the help during the preparation of the observations and their execution.
We are grateful to the SPHERE consortium for providing us with the most recent measurements of the instrument platescale and True North. We thank Aurélia Leclerc from IPAG for fruitful discussions about the robustness of the Gaia-eDR3 and HIPPARCOS proper motion determination. We acknowledge support in France from the French National Research Agency (ANR) through project grants ANR-14-CE33-0018 and ANR-20-CE31-0012, the CNRS-D2P PICS grant, and the Programmes Nationaux de Planetologie et de Physique Stellaire (PNP \& PNPS). This project has received funding from the European Research Council (ERC) under the European Union's Horizon 2020 research and innovation programme (COBREX; grant agreement 885593). SDe, RGr, CLa, and Dme acknowledge financial support from the ASI-INAF agreement n.2018-16-HH.0 and from PRIN-INAF 2019 "Planetary systems at young ages (PLATEA)". C.P. acknowledges funding from the Australian Research Council via FT170100040 and DP180104235. J.\,O. acknowledges support by ANID, -- Millennium Science Initiative Program -- NCN19\_171, from the Universidad de Valpara\'iso, and from Fondecyt (grant 1180395). 

This work has made use of the SPHERE Data Centre, jointly operated by OSUG/IPAG (Grenoble), PYTHEAS/LAM/CeSAM (Marseille), OCA/Lagrange (Nice), Observatoire de Paris/LESIA (Paris), and Observatoire de Lyon/CRAL, and supported by a grant from Labex OSUG@2020 (Investissements d’avenir – ANR10 LABX56).

This publication makes use of VOSA, developed under the Spanish Virtual Observatory project supported by the Spanish MINECO through grant AyA2017-84089. VOSA has been partially updated by using funding from the European Union's Horizon 2020 Research and Innovation Programme, under Grant Agreement nº 776403 (EXOPLANETS-A).
\end{acknowledgements}

\bibliographystyle{aa}
\bibliography{HIP77432_Bonnefoy_v0}

\begin{thebibliography}{86}
\expandafter\ifx\csname natexlab\endcsname\relax\def\natexlab#1{#1}\fi

\bibitem[{{Adam} {et~al.}(2021){Adam}, {Olofsson}, {van Holstein}, {Bayo},
  {Milli}, {Boccaletti}, {Kral}, {Ginski}, {Henning}, {Montesinos}, {Pawellek},
  {Zurlo}, {Langlois}, {Delboulbe}, {Pavlov}, {Ramos}, {Weber}, {Wildi},
  {Rigal}, \& {Sauvage}}]{2021arXiv210706316A}
{Adam}, C., {Olofsson}, J., {van Holstein}, R.~G., {et~al.} 2021, arXiv
  e-prints, arXiv:2107.06316

\bibitem[{{Arnold} {et~al.}(2019){Arnold}, {Weinberger}, {Videen}, \&
  {Zubko}}]{Arnold2019}
{Arnold}, J.~A., {Weinberger}, A.~J., {Videen}, G., \& {Zubko}, E.~S. 2019,
  arXiv e-prints [\eprint[arXiv]{1902.10183}]

\bibitem[{{Augereau} {et~al.}(1999){Augereau}, {Lagrange}, {Mouillet},
  {Papaloizou}, \& {Grorod}}]{Augereau1999}
{Augereau}, J.~C., {Lagrange}, A.~M., {Mouillet}, D., {Papaloizou}, J.~C.~B.,
  \& {Grorod}, P.~A. 1999, \aap, 348, 557

\bibitem[{{Ballering} {et~al.}(2017){Ballering}, {Rieke}, {Su}, \&
  {G{\'a}sp{\'a}r}}]{Ballering2017}
{Ballering}, N.~P., {Rieke}, G.~H., {Su}, K. Y.~L., \& {G{\'a}sp{\'a}r}, A.
  2017, \apj, 845, 120

\bibitem[{{Baraffe} {et~al.}(2003){Baraffe}, {Chabrier}, {Barman}, {Allard}, \&
  {Hauschildt}}]{2003A&A...402..701B}
{Baraffe}, I., {Chabrier}, G., {Barman}, T.~S., {Allard}, F., \& {Hauschildt},
  P.~H. 2003, \aap, 402, 701

\bibitem[{{Bayo} {et~al.}(2008){Bayo}, {Rodrigo}, {Barrado Y Navascu{\'e}s},
  {Solano}, {Guti{\'e}rrez}, {Morales-Calder{\'o}n}, \&
  {Allard}}]{2008A&A...492..277B}
{Bayo}, A., {Rodrigo}, C., {Barrado Y Navascu{\'e}s}, D., {et~al.} 2008, \aap,
  492, 277

\bibitem[{{Beuzit} {et~al.}(2019){Beuzit}, {Vigan}, {Mouillet}, {Dohlen},
  {Gratton}, {Boccaletti}, {Sauvage}, {Schmid}, {Langlois}, {Petit},
  {Baruffolo}, {Feldt}, {Milli}, {Wahhaj}, {Abe}, {Anselmi}, {Antichi},
  {Barette}, {Baudrand}, {Baudoz}, {Bazzon}, {Bernardi}, {Blanchard}, {Brast},
  {Bruno}, {Buey}, {Carbillet}, {Carle}, {Cascone}, {Chapron}, {Charton},
  {Chauvin}, {Claudi}, {Costille}, {De Caprio}, {de Boer}, {Delboulb{\'e}},
  {Desidera}, {Dominik}, {Downing}, {Dupuis}, {Fabron}, {Fantinel}, {Farisato},
  {Feautrier}, {Fedrigo}, {Fusco}, {Gigan}, {Ginski}, {Girard}, {Giro},
  {Gisler}, {Gluck}, {Gry}, {Henning}, {Hubin}, {Hugot}, {Incorvaia}, {Jaquet},
  {Kasper}, {Lagadec}, {Lagrange}, {Le Coroller}, {Le Mignant}, {Le Ruyet},
  {Lessio}, {Lizon}, {Llored}, {Lundin}, {Madec}, {Magnard}, {Marteaud},
  {Martinez}, {Maurel}, {M{\'e}nard}, {Mesa}, {M{\"o}ller-Nilsson}, {Moulin},
  {Moutou}, {Orign{\'e}}, {Parisot}, {Pavlov}, {Perret}, {Pragt}, {Puget},
  {Rabou}, {Ramos}, {Reess}, {Rigal}, {Rochat}, {Roelfsema}, {Rousset}, {Roux},
  {Saisse}, {Salasnich}, {Santambrogio}, {Scuderi}, {Segransan}, {Sevin},
  {Siebenmorgen}, {Soenke}, {Stadler}, {Suarez}, {Tiph{\`e}ne}, {Turatto},
  {Udry}, {Vakili}, {Waters}, {Weber}, {Wildi}, {Zins}, \&
  {Zurlo}}]{2019A&A...631A.155B}
{Beuzit}, J.~L., {Vigan}, A., {Mouillet}, D., {et~al.} 2019, \aap, 631, A155

\bibitem[{{Boccaletti} {et~al.}(2020){Boccaletti}, {Chauvin}, {Mouillet},
  {Absil}, {Allard}, {Antoniucci}, {Augereau}, {Barge}, {Baruffolo}, {Baudino},
  {Baudoz}, {Beaulieu}, {Benisty}, {Beuzit}, {Bianco}, {Biller}, {Bonavita},
  {Bonnefoy}, {Bos}, {Bouret}, {Brandner}, {Buchschache}, {Carry},
  {Cantalloube}, {Cascone}, {Carlotti}, {Charnay}, {Chiavassa}, {Choquet},
  {Clenet}, {Crida}, {De Boer}, {De Caprio}, {Desidera}, {Desert}, {Delisle},
  {Delorme}, {Dohlen}, {Doelman}, {Dominik}, {Orazi}, {Dougados}, {Doute},
  {Fedele}, {Feldt}, {Ferreira}, {Fontanive}, {Fusco}, {Galicher}, {Garufi},
  {Gendron}, {Ghedina}, {Ginski}, {Gonzalez}, {Gratadour}, {Gratton},
  {Guillot}, {Haffert}, {Hagelberg}, {Henning}, {Huby}, {Janson}, {Kamp},
  {Keller}, {Kenworthy}, {Kervella}, {Kral}, {Kuhn}, {Lagadec}, {Laibe},
  {Langlois}, {Lagrange}, {Launhardt}, {Leboulleux}, {Le Coroller}, {Li Causi},
  {Loupias}, {Maire}, {Marleau}, {Martinache}, {Martinez}, {Mary}, {Mattioli},
  {Mazoyer}, {Meheut}, {Menard}, {Mesa}, {Meunier}, {Miguel}, {Milli}, {Min},
  {Molliere}, {Mordasini}, {Moretto}, {Mugnier}, {Muro Arena}, {Nardetto},
  {Diaye}, {Nesvadba}, {Pedichini}, {Pinilla}, {Por}, {Potier}, {Quanz},
  {Rameau}, {Roelfsema}, {Rouan}, {Rigliaco}, {Salasnich}, {Samland},
  {Sauvage}, {Schmid}, {Segransan}, {Snellen}, {Snik}, {Soulez}, {Stadler},
  {Stam}, {Tallon}, {Thebault}, {Thiebaut}, {Tschudi}, {Udry}, {van Holstein},
  {Vernazza}, {Vidal}, {Vigan}, {Waters}, {Wildi}, {Willson}, {Zanutta},
  {Zavagno}, \& {Zurlo}}]{2020arXiv200305714B}
{Boccaletti}, A., {Chauvin}, G., {Mouillet}, D., {et~al.} 2020, arXiv e-prints,
  arXiv:2003.05714

\bibitem[{{Bohn} {et~al.}(2019){Bohn}, {Kenworthy}, {Ginski}, {Benisty}, {de
  Boer}, {Keller}, {Mamajek}, {Meshkat}, {Muro-Arena}, {Pecaut}, {Snik},
  {Wolff}, \& {Reggiani}}]{2019A&A...624A..87B}
{Bohn}, A.~J., {Kenworthy}, M.~A., {Ginski}, C., {et~al.} 2019, \aap, 624, A87

\bibitem[{{Bonnefoy} {et~al.}(2017){Bonnefoy}, {Milli}, {M{\'e}nard}, {Vigan},
  {Lagrange}, {Delorme}, {Boccaletti}, {Lazzoni}, {Galicher}, {Desidera},
  {Chauvin}, {Augereau}, {Mouillet}, {Pinte}, {van der Plas}, {Gratton},
  {Beust}, \& {Beuzit}}]{Bonnefoy2017}
{Bonnefoy}, M., {Milli}, J., {M{\'e}nard}, F., {et~al.} 2017, \aap, 597, L7

\bibitem[{{Cantalloube} {et~al.}(2020){Cantalloube}, {Farley}, {Milli},
  {Bharmal}, {Brandner}, {Correia}, {Dohlen}, {Henning}, {Osborn}, {Por},
  {Su{\'a}rez Valles}, \& {Vigan}}]{Cantalloube2020}
{Cantalloube}, F., {Farley}, O.~J.~D., {Milli}, J., {et~al.} 2020, \aap, 638,
  A98

\bibitem[{{Chen} {et~al.}(2014){Chen}, {Mittal}, {Kuchner}, {Forrest}, {Lisse},
  {Manoj}, {Sargent}, \& {Watson}}]{2014ApJS..211...25C}
{Chen}, C.~H., {Mittal}, T., {Kuchner}, M., {et~al.} 2014, \apjs, 211, 25

\bibitem[{{Claudi} {et~al.}(2008){Claudi}, {Turatto}, {Gratton}, {Antichi},
  {Bonavita}, {Bruno}, {Cascone}, {De Caprio}, {Desidera}, {Giro}, {Mesa},
  {Scuderi}, {Dohlen}, {Beuzit}, \& {Puget}}]{2008SPIE.7014E..3EC}
{Claudi}, R.~U., {Turatto}, M., {Gratton}, R.~G., {et~al.} 2008, in Society of
  Photo-Optical Instrumentation Engineers (SPIE) Conference Series, Vol. 7014,
  Society of Photo-Optical Instrumentation Engineers (SPIE) Conference Series,
  3

\bibitem[{{Currie} {et~al.}(2015){Currie}, {Lisse}, {Kuchner}, {Madhusudhan},
  {Kenyon}, {Thalmann}, {Carson}, \& {Debes}}]{Currie2015}
{Currie}, T., {Lisse}, C.~M., {Kuchner}, M., {et~al.} 2015, \apjl, 807, L7

\bibitem[{{Cutri} {et~al.}(2003){Cutri}, {Skrutskie}, {van Dyk}, {Beichman},
  {Carpenter}, {Chester}, {Cambresy}, {Evans}, {Fowler}, {Gizis}, {Howard},
  {Huchra}, {Jarrett}, {Kopan}, {Kirkpatrick}, {Light}, {Marsh}, {McCallon},
  {Schneider}, {Stiening}, {Sykes}, {Weinberg}, {Wheaton}, {Wheelock}, \&
  {Zacarias}}]{2003yCat.2246....0C}
{Cutri}, R.~M., {Skrutskie}, M.~F., {van Dyk}, S., {et~al.} 2003, VizieR Online
  Data Catalog, II/246

\bibitem[{{de Zeeuw} {et~al.}(1999){de Zeeuw}, {Hoogerwerf}, {de Bruijne},
  {Brown}, \& {Blaauw}}]{1999AJ....117..354D}
{de Zeeuw}, P.~T., {Hoogerwerf}, R., {de Bruijne}, J.~H.~J., {Brown}, A.~G.~A.,
  \& {Blaauw}, A. 1999, \aj, 117, 354

\bibitem[{{Delorme} {et~al.}(2017){Delorme}, {Meunier}, {Albert}, {Lagadec},
  {Le Coroller}, {Galicher}, {Mouillet}, {Boccaletti}, {Mesa}, {Meunier},
  {Beuzit}, {Lagrange}, {Chauvin}, {Sapone}, {Langlois}, {Maire},
  {Montarg{\`e}s}, {Gratton}, {Vigan}, \& {Surace}}]{Delorme2017_DC}
{Delorme}, P., {Meunier}, N., {Albert}, D., {et~al.} 2017, in SF2A-2017:
  Proceedings of the Annual meeting of the French Society of Astronomy and
  Astrophysics, ed. C.~{Reyl{\'e}}, P.~{Di Matteo}, F.~{Herpin}, E.~{Lagadec},
  A.~{Lan{\c{c}}on}, Z.~{Meliani}, \& F.~{Royer}, Di

\bibitem[{{Dohlen} {et~al.}(2008){Dohlen}, {Langlois}, {Saisse}, {Hill},
  {Origne}, {Jacquet}, {Fabron}, {Blanc}, {Llored}, {Carle}, {Moutou}, {Vigan},
  {Boccaletti}, {Carbillet}, {Mouillet}, \& {Beuzit}}]{2008SPIE.7014E..3LD}
{Dohlen}, K., {Langlois}, M., {Saisse}, M., {et~al.} 2008, in Society of
  Photo-Optical Instrumentation Engineers (SPIE) Conference Series, Vol. 7014,
  Society of Photo-Optical Instrumentation Engineers (SPIE) Conference Series,
  3

\bibitem[{{Dohnanyi}(1969)}]{Dohnanyi1969}
{Dohnanyi}, J.~S. 1969, \jgr, 74, 2531

\bibitem[{{Draine} \& {Lee}(1984)}]{Draine1984}
{Draine}, B.~T. \& {Lee}, H.~M. 1984, \apj, 285, 89

\bibitem[{{Draper} {et~al.}(2016){Draper}, {Duch{\^e}ne}, {Millar-Blanchaer},
  {Matthews}, {Wang}, {Kalas}, {Graham}, {Padgett}, {Ammons}, {Bulger}, {Chen},
  {Chilcote}, {Doyon}, {Fitzgerald}, {Follette}, {Gerard}, {Greenbaum},
  {Hibon}, {Hinkley}, {Macintosh}, {Ingraham}, {Lafreni{\`e}re}, {Marchis},
  {Marois}, {Nielsen}, {Oppenheimer}, {Patel}, {Patience}, {Perrin}, {Pueyo},
  {Rajan}, {Rameau}, {Sivaramakrishnan}, {Vega}, {Ward-Duong}, \&
  {Wolff}}]{2016arXiv160502771D}
{Draper}, Z.~H., {Duch{\^e}ne}, G., {Millar-Blanchaer}, M.~A., {et~al.} 2016,
  ArXiv e-prints [\eprint[arXiv]{1605.02771}]

\bibitem[{{Ducourant} {et~al.}(2014){Ducourant}, {Teixeira}, {Galli}, {Le
  Campion}, {Krone-Martins}, {Zuckerman}, {Chauvin}, \&
  {Song}}]{2014A&A...563A.121D}
{Ducourant}, C., {Teixeira}, R., {Galli}, P.~A.~B., {et~al.} 2014, \aap, 563,
  A121

\bibitem[{{Engler} {et~al.}(2020){Engler}, {Lazzoni}, {Gratton}, {Milli},
  {Schmid}, {Chauvin}, {Kral}, {Pawellek}, {Th{\'e}bault}, {Boccaletti},
  {Bonnefoy}, {Brown}, {Buey}, {Cantalloube}, {Carle}, {Cheetham}, {Desidera},
  {Feldt}, {Ginski}, {Gisler}, {Henning}, {Hunziker}, {Lagrange}, {Langlois},
  {Mesa}, {Meyer}, {Moeller-Nilsson}, {Olofsson}, {Petit}, {Petrus}, {Quanz},
  {Rickman}, {Stadler}, {Stolker}, {Vigan}, {Wildi}, \&
  {Zurlo}}]{2020A&A...635A..19E}
{Engler}, N., {Lazzoni}, C., {Gratton}, R., {et~al.} 2020, \aap, 635, A19

\bibitem[{{Engler} {et~al.}(2017){Engler}, {Schmid}, {Thalmann}, {Boccaletti},
  {Bazzon}, {Baruffolo}, {Beuzit}, {Claudi}, {Costille}, {Desidera}, {Dohlen},
  {Dominik}, {Feldt}, {Fusco}, {Ginski}, {Gisler}, {Girard}, {Gratton},
  {Henning}, {Hubin}, {Janson}, {Kasper}, {Kral}, {Langlois}, {Lagadec},
  {M{\'e}nard}, {Meyer}, {Milli}, {Mouillet}, {Olofsson}, {Pavlov}, {Pragt},
  {Puget}, {Quanz}, {Roelfsema}, {Salasnich}, {Siebenmorgen}, {Sissa},
  {Suarez}, {Szulagyi}, {Turatto}, {Udry}, \& {Wildi}}]{2017A&A...607A..90E}
{Engler}, N., {Schmid}, H.~M., {Thalmann}, C., {et~al.} 2017, \aap, 607, A90

\bibitem[{{Esposito} {et~al.}(2020){Esposito}, {Kalas}, {Fitzgerald},
  {Millar-Blanchaer}, {Duch{\^e}ne}, {Patience}, {Hom}, {Perrin}, {De Rosa},
  {Chiang}, {Czekala}, {Macintosh}, {Graham}, {Ansdell}, {Arriaga}, {Bruzzone},
  {Bulger}, {Chen}, {Cotten}, {Dong}, {Draper}, {Follette}, {Hung}, {Lopez},
  {Matthews}, {Mazoyer}, {Metchev}, {Rameau}, {Ren}, {Rice}, {Song}, {Stahl},
  {Wang}, {Wolff}, {Zuckerman}, {Ammons}, {Bailey}, {Barman}, {Chilcote},
  {Doyon}, {Gerard}, {Goodsell}, {Greenbaum}, {Hibon}, {Hinkley}, {Ingraham},
  {Konopacky}, {Maire}, {Marchis}, {Marley}, {Marois}, {Nielsen},
  {Oppenheimer}, {Palmer}, {Poyneer}, {Pueyo}, {Rajan}, {Rantakyr{\"o}},
  {Ruffio}, {Savransky}, {Schneider}, {Sivaramakrishnan}, {Soummer}, {Thomas},
  \& {Ward-Duong}}]{Esposito2020}
{Esposito}, T.~M., {Kalas}, P., {Fitzgerald}, M.~P., {et~al.} 2020, \aj, 160,
  24

\bibitem[{{Feldt, M.} {et~al.}(2017){Feldt, M.}, {Olofsson, J.}, {Boccaletti,
  A.}, {Maire, A. L.}, {Milli, J.}, {Vigan, A.}, {Langlois, M.}, {Henning,
  Th.}, {Moor, A.}, {Bonnefoy, M.}, {Wahhaj, Z.}, {Desidera, S.}, {Gratton,
  R.}, {K\'osp\'al, \'A.}, {Abraham, P.}, {Menard, F.}, {Chauvin, G.},
  {Lagrange, A. M.}, {Mesa, D.}, {Salter, G.}, {Buenzli, E.}, {Lannier, J.},
  {Perrot, C.}, {Peretti, S.}, \& {Sissa, E.}}]{refId0}
{Feldt, M.}, {Olofsson, J.}, {Boccaletti, A.}, {et~al.} 2017, A\&A, 601, A7

\bibitem[{{Gagn{\'e}} {et~al.}(2018){Gagn{\'e}}, {Mamajek}, {Malo}, {Riedel},
  {Rodriguez}, {Lafreni{\`e}re}, {Faherty}, {Roy-Loubier}, {Pueyo}, {Robin}, \&
  {Doyon}}]{2018ApJ...856...23G}
{Gagn{\'e}}, J., {Mamajek}, E.~E., {Malo}, L., {et~al.} 2018, \apj, 856, 23

\bibitem[{{Gaia Collaboration}(2020)}]{2020yCat.1350....0G}
{Gaia Collaboration}. 2020, VizieR Online Data Catalog, I/350

\bibitem[{{Galicher} {et~al.}(2018){Galicher}, {Boccaletti}, {Mesa}, {Delorme},
  {Gratton}, {Langlois}, {Lagrange}, {Maire}, {Le Coroller}, {Chauvin},
  {Biller}, {Cantalloube}, {Janson}, {Lagadec}, {Meunier}, {Vigan},
  {Hagelberg}, {Bonnefoy}, {Zurlo}, {Rocha}, {Maurel}, {Jaquet}, {Buey}, \&
  {Weber}}]{Galicher2018_DC}
{Galicher}, R., {Boccaletti}, A., {Mesa}, D., {et~al.} 2018, \aap, 615, A92

\bibitem[{{Galland} {et~al.}(2005){Galland}, {Lagrange}, {Udry}, {Chelli},
  {Pepe}, {Queloz}, {Beuzit}, \& {Mayor}}]{2005A&A...443..337G}
{Galland}, F., {Lagrange}, A.~M., {Udry}, S., {et~al.} 2005, \aap, 443, 337

\bibitem[{{Gibbs} {et~al.}(2019){Gibbs}, {Wagner}, {Apai}, {Mo{\'o}r},
  {Currie}, {Bonnefoy}, {Langlois}, \& {Lisse}}]{2019AJ....157...39G}
{Gibbs}, A., {Wagner}, K., {Apai}, D., {et~al.} 2019, \aj, 157, 39

\bibitem[{{Gomes}(2021)}]{2021Icar..35714121G}
{Gomes}, R. 2021, \icarus, 357, 114121

\bibitem[{{Gomez Gonzalez} {et~al.}(2017){Gomez Gonzalez}, {Wertz}, {Absil},
  {Christiaens}, {Defr{\`e}re}, {Mawet}, {Milli}, {Absil}, {Van Droogenbroeck},
  {Cantalloube}, {Hinz}, {Skemer}, {Karlsson}, \& {Surdej}}]{Gomez2017}
{Gomez Gonzalez}, C.~A., {Wertz}, O., {Absil}, O., {et~al.} 2017, \aj, 154, 7

\bibitem[{{Hom} {et~al.}(2020){Hom}, {Patience}, {Esposito}, {Duch{\^e}ne},
  {Worthen}, {Kalas}, {Jang-Condell}, {Saboi}, {Arriaga}, {Mazoyer}, {Wolff},
  {Millar-Blanchaer}, {Fitzgerald}, {Perrin}, {Chen}, {Macintosh}, {Matthews},
  {Wang}, {Graham}, {Marchis}, {Ammons}, {Bailey}, {Barman}, {Bulger},
  {Chilcote}, {Cotten}, {De Rosa}, {Doyon}, {Follette}, {Goodsell},
  {Greenbaum}, {Hibon}, {Ingraham}, {Konopacky}, {Larkin}, {Maire}, {Marley},
  {Marois}, {Matthews}, {Metchev}, {Nielsen}, {Oppenheimer}, {Palmer},
  {Poyneer}, {Pueyo}, {Rajan}, {Rameau}, {Rantakyr{\"o}}, {Ren}, {Savransky},
  {Schneider}, {Sivaramakrishnan}, {Song}, {Soummer}, {Tallis}, {Thomas},
  {Wallace}, {Ward-Duong}, {Wiktorowicz}, \& {Zuckerman}}]{Hom2020}
{Hom}, J., {Patience}, J., {Esposito}, T.~M., {et~al.} 2020, \aj, 159, 31

\bibitem[{{Jang-Condell} {et~al.}(2015){Jang-Condell}, {Chen}, {Mittal},
  {Manoj}, {Watson}, {Lisse}, {Nesvold}, \& {Kuchner}}]{2015ApJ...808..167J}
{Jang-Condell}, H., {Chen}, C.~H., {Mittal}, T., {et~al.} 2015, \apj, 808, 167

\bibitem[{{Janson} {et~al.}(2013){Janson}, {Lafreni{\`e}re}, {Jayawardhana},
  {Bonavita}, {Girard}, {Brandeker}, \& {Gizis}}]{2013ApJ...773..170J}
{Janson}, M., {Lafreni{\`e}re}, D., {Jayawardhana}, R., {et~al.} 2013, \apj,
  773, 170

\bibitem[{{Kalas} {et~al.}(2005){Kalas}, {Graham}, \&
  {Clampin}}]{2005Natur.435.1067K}
{Kalas}, P., {Graham}, J.~R., \& {Clampin}, M. 2005, \nat, 435, 1067

\bibitem[{{Kalas} {et~al.}(2013){Kalas}, {Graham}, {Fitzgerald}, \&
  {Clampin}}]{2013ApJ...775...56K}
{Kalas}, P., {Graham}, J.~R., {Fitzgerald}, M.~P., \& {Clampin}, M. 2013, \apj,
  775, 56

\bibitem[{{Kalas} {et~al.}(2015){Kalas}, {Rajan}, {Wang}, {Millar-Blanchaer},
  {Duchene}, {Chen}, {Fitzgerald}, {Dong}, {Graham}, {Patience}, {Macintosh},
  {Murray-Clay}, {Matthews}, {Rameau}, {Marois}, {Chilcote}, {De Rosa},
  {Doyon}, {Draper}, {Lawler}, {Ammons}, {Arriaga}, {Bulger}, {Cotten},
  {Follette}, {Goodsell}, {Greenbaum}, {Hibon}, {Hinkley}, {Hung}, {Ingraham},
  {Konapacky}, {Lafreniere}, {Larkin}, {Long}, {Maire}, {Marchis}, {Metchev},
  {Morzinski}, {Nielsen}, {Oppenheimer}, {Perrin}, {Pueyo}, {Rantakyr{\"o}},
  {Ruffio}, {Saddlemyer}, {Savransky}, {Schneider}, {Sivaramakrishnan},
  {Soummer}, {Song}, {Thomas}, {Vasisht}, {Ward-Duong}, {Wiktorowicz}, \&
  {Wolff}}]{2015ApJ...814...32K}
{Kalas}, P.~G., {Rajan}, A., {Wang}, J.~J., {et~al.} 2015, \apj, 814, 32

\bibitem[{{Kasper} {et~al.}(2015){Kasper}, {Apai}, {Wagner}, \&
  {Robberto}}]{2015ApJ...812L..33K}
{Kasper}, M., {Apai}, D., {Wagner}, K., \& {Robberto}, M. 2015, \apjl, 812, L33

\bibitem[{{Kennedy} \& {Wyatt}(2014)}]{2014MNRAS.444.3164K}
{Kennedy}, G.~M. \& {Wyatt}, M.~C. 2014, \mnras, 444, 3164

\bibitem[{{Kervella} {et~al.}(2021){Kervella}, {Arenou}, \&
  {Th{\'e}venin}}]{2021arXiv210910912K}
{Kervella}, P., {Arenou}, F., \& {Th{\'e}venin}, F. 2021, arXiv e-prints,
  arXiv:2109.10912

\bibitem[{{Kral} {et~al.}(2020){Kral}, {Matr{\`a}}, {Kennedy}, {Marino}, \&
  {Wyatt}}]{2020MNRAS.497.2811K}
{Kral}, Q., {Matr{\`a}}, L., {Kennedy}, G.~M., {Marino}, S., \& {Wyatt}, M.~C.
  2020, \mnras, 497, 2811

\bibitem[{{Lafreni{\`e}re} {et~al.}(2007){Lafreni{\`e}re}, {Marois}, {Doyon},
  {Nadeau}, \& {Artigau}}]{2007ApJ...660..770L}
{Lafreni{\`e}re}, D., {Marois}, C., {Doyon}, R., {Nadeau}, D., \& {Artigau},
  {\'E}. 2007, \apj, 660, 770

\bibitem[{{Lagrange} {et~al.}(2016){Lagrange}, {Langlois}, {Gratton}, {Maire},
  {Milli}, {Olofsson}, {Vigan}, {Bailey}, {Mesa}, {Chauvin}, {Boccaletti},
  {Galicher}, {Girard}, {Bonnefoy}, {Samland}, {Menard}, {Henning},
  {Kenworthy}, {Thalmann}, {Beust}, {Beuzit}, {Brandner}, {Buenzli},
  {Cheetham}, {Janson}, {le Coroller}, {Lannier}, {Mouillet}, {Peretti},
  {Perrot}, {Salter}, {Sissa}, {Wahhaj}, {Abe}, {Desidera}, {Feldt}, {Madec},
  {Perret}, {Petit}, {Rabou}, {Soenke}, \& {Weber}}]{2016A&A...586L...8L}
{Lagrange}, A.-M., {Langlois}, M., {Gratton}, R., {et~al.} 2016, \aap, 586, L8

\bibitem[{{Lagrange} {et~al.}(2012){Lagrange}, {Milli}, {Boccaletti}, {Lacour},
  {Thebault}, {Chauvin}, {Mouillet}, {Augereau}, {Bonnefoy}, {Ehrenreich}, \&
  {Kral}}]{2012A&A...546A..38L}
{Lagrange}, A.~M., {Milli}, J., {Boccaletti}, A., {et~al.} 2012, \aap, 546, A38

\bibitem[{{Lannier} {et~al.}(2017){Lannier}, {Lagrange}, {Bonavita},
  {Borgniet}, {Delorme}, {Meunier}, {Desidera}, {Messina}, {Chauvin}, \&
  {Keppler}}]{2017A&A...603A..54L}
{Lannier}, J., {Lagrange}, A.~M., {Bonavita}, M., {et~al.} 2017, \aap, 603, A54

\bibitem[{{Lee} \& {Chiang}(2016)}]{Lee2016}
{Lee}, E.~J. \& {Chiang}, E. 2016, \apj, 827, 125

\bibitem[{{Li} \& {Greenberg}(1998)}]{Li1998}
{Li}, A. \& {Greenberg}, J.~M. 1998, \aap, 331, 291

\bibitem[{{Lieman-Sifry} {et~al.}(2016){Lieman-Sifry}, {Hughes}, {Carpenter},
  {Gorti}, {Hales}, \& {Flaherty}}]{2016ApJ...828...25L}
{Lieman-Sifry}, J., {Hughes}, A.~M., {Carpenter}, J.~M., {et~al.} 2016, \apj,
  828, 25

\bibitem[{{MacGregor} {et~al.}(2017){MacGregor}, {Matr{\`a}}, {Kalas},
  {Wilner}, {Pan}, {Kennedy}, {Wyatt}, {Duchene}, {Hughes}, {Rieke}, {Clampin},
  {Fitzgerald}, {Graham}, {Holland}, {Pani{\'c}}, {Shannon}, \&
  {Su}}]{2017ApJ...842....8M}
{MacGregor}, M.~A., {Matr{\`a}}, L., {Kalas}, P., {et~al.} 2017, \apj, 842, 8

\bibitem[{{Macintosh} {et~al.}(2015){Macintosh}, {Graham}, {Barman}, {De Rosa},
  {Konopacky}, {Marley}, {Marois}, {Nielsen}, {Pueyo}, {Rajan}, {Rameau},
  {Saumon}, {Wang}, {Patience}, {Ammons}, {Arriaga}, {Artigau}, {Beckwith},
  {Brewster}, {Bruzzone}, {Bulger}, {Burningham}, {Burrows}, {Chen}, {Chiang},
  {Chilcote}, {Dawson}, {Dong}, {Doyon}, {Draper}, {Duch{\^e}ne}, {Esposito},
  {Fabrycky}, {Fitzgerald}, {Follette}, {Fortney}, {Gerard}, {Goodsell},
  {Greenbaum}, {Hibon}, {Hinkley}, {Cotten}, {Hung}, {Ingraham},
  {Johnson-Groh}, {Kalas}, {Lafreniere}, {Larkin}, {Lee}, {Line}, {Long},
  {Maire}, {Marchis}, {Matthews}, {Max}, {Metchev}, {Millar-Blanchaer},
  {Mittal}, {Morley}, {Morzinski}, {Murray-Clay}, {Oppenheimer}, {Palmer},
  {Patel}, {Perrin}, {Poyneer}, {Rafikov}, {Rantakyr{\"o}}, {Rice}, {Rojo},
  {Rudy}, {Ruffio}, {Ruiz}, {Sadakuni}, {Saddlemyer}, {Salama}, {Savransky},
  {Schneider}, {Sivaramakrishnan}, {Song}, {Soummer}, {Thomas}, {Vasisht},
  {Wallace}, {Ward-Duong}, {Wiktorowicz}, {Wolff}, \&
  {Zuckerman}}]{2015Sci...350...64M}
{Macintosh}, B., {Graham}, J.~R., {Barman}, T., {et~al.} 2015, Science, 350, 64

\bibitem[{{Macintosh} {et~al.}(2008){Macintosh}, {Graham}, {Palmer}, {Doyon},
  {Dunn}, {Gavel}, {Larkin}, {Oppenheimer}, {Saddlemyer}, {Sivaramakrishnan},
  {Wallace}, {Bauman}, {Erickson}, {Marois}, {Poyneer}, \&
  {Soummer}}]{2008SPIE.7015E..18M}
{Macintosh}, B.~A., {Graham}, J.~R., {Palmer}, D.~W., {et~al.} 2008, in
  \procspie, Vol. 7015, Adaptive Optics Systems, 701518

\bibitem[{{Maire} {et~al.}(2016){Maire}, {Bonnefoy}, {Ginski}, {Vigan},
  {Messina}, {Mesa}, {Galicher}, {Gratton}, {Desidera}, {Kopytova}, {Millward},
  {Thalmann}, {Claudi}, {Ehrenreich}, {Zurlo}, {Chauvin}, {Antichi},
  {Baruffolo}, {Bazzon}, {Beuzit}, {Blanchard}, {Boccaletti}, {de Boer},
  {Carle}, {Cascone}, {Costille}, {De Caprio}, {Delboulb{\'e}}, {Dohlen},
  {Dominik}, {Feldt}, {Fusco}, {Girard}, {Giro}, {Gisler}, {Gluck}, {Gry},
  {Henning}, {Hubin}, {Hugot}, {Jaquet}, {Kasper}, {Lagrange}, {Langlois}, {Le
  Mignant}, {Llored}, {Madec}, {Martinez}, {Mawet}, {Milli},
  {M{\"o}ller-Nilsson}, {Mouillet}, {Moulin}, {Moutou}, {Orign{\'e}}, {Pavlov},
  {Petit}, {Pragt}, {Puget}, {Ramos}, {Rochat}, {Roelfsema}, {Salasnich},
  {Sauvage}, {Schmid}, {Turatto}, {Udry}, {Vakili}, {Wahhaj}, {Weber}, \&
  {Wildi}}]{2016A&A...587A..56M}
{Maire}, A.-L., {Bonnefoy}, M., {Ginski}, C., {et~al.} 2016, \aap, 587, A56

\bibitem[{{Mamajek} {et~al.}(2002){Mamajek}, {Meyer}, \&
  {Liebert}}]{2002AJ....124.1670M}
{Mamajek}, E.~E., {Meyer}, M.~R., \& {Liebert}, J. 2002, \aj, 124, 1670

\bibitem[{{Marois} {et~al.}(2006){Marois}, {Lafreni{\`e}re}, {Doyon},
  {Macintosh}, \& {Nadeau}}]{2006ApJ...641..556M}
{Marois}, C., {Lafreni{\`e}re}, D., {Doyon}, R., {Macintosh}, B., \& {Nadeau},
  D. 2006, \apj, 641, 556

\bibitem[{{Marois} {et~al.}(2008){Marois}, {Macintosh}, {Barman}, {Zuckerman},
  {Song}, {Patience}, {Lafreni{\`e}re}, \& {Doyon}}]{2008Sci...322.1348M}
{Marois}, C., {Macintosh}, B., {Barman}, T., {et~al.} 2008, Science, 322, 1348

\bibitem[{{Marois} {et~al.}(2010){Marois}, {Zuckerman}, {Konopacky},
  {Macintosh}, \& {Barman}}]{2010Natur.468.1080M}
{Marois}, C., {Zuckerman}, B., {Konopacky}, Q.~M., {Macintosh}, B., \&
  {Barman}, T. 2010, \nat, 468, 1080

\bibitem[{{Matr{\`a}} {et~al.}(2018){Matr{\`a}}, {Marino}, {Kennedy}, {Wyatt},
  {{\"O}berg}, \& {Wilner}}]{2018ApJ...859...72M}
{Matr{\`a}}, L., {Marino}, S., {Kennedy}, G.~M., {et~al.} 2018, \apj, 859, 72

\bibitem[{{Matthews} {et~al.}(2017){Matthews}, {Hinkley}, {Vigan}, {Kennedy},
  {Rizzuto}, {Stapelfeldt}, {Mawet}, {Booth}, {Chen}, \&
  {Jang-Condell}}]{2017ApJ...843L..12M}
{Matthews}, E., {Hinkley}, S., {Vigan}, A., {et~al.} 2017, \apjl, 843, L12

\bibitem[{{Mawet} {et~al.}(2014){Mawet}, {Milli}, {Wahhaj}, {Pelat}, {Absil},
  {Delacroix}, {Boccaletti}, {Kasper}, {Kenworthy}, {Marois}, {Mennesson}, \&
  {Pueyo}}]{Mawet2014}
{Mawet}, D., {Milli}, J., {Wahhaj}, Z., {et~al.} 2014, \apj, 792, 97

\bibitem[{{Mayor} {et~al.}(2003){Mayor}, {Pepe}, {Queloz}, {Bouchy},
  {Rupprecht}, {Lo Curto}, {Avila}, {Benz}, {Bertaux}, {Bonfils}, {Dall},
  {Dekker}, {Delabre}, {Eckert}, {Fleury}, {Gilliotte}, {Gojak}, {Guzman},
  {Kohler}, {Lizon}, {Longinotti}, {Lovis}, {Megevand}, {Pasquini}, {Reyes},
  {Sivan}, {Sosnowska}, {Soto}, {Udry}, {van Kesteren}, {Weber}, \&
  {Weilenmann}}]{2003Msngr.114...20M}
{Mayor}, M., {Pepe}, F., {Queloz}, D., {et~al.} 2003, The Messenger, 114, 20

\bibitem[{{Mesa} {et~al.}(2015){Mesa}, {Gratton}, {Zurlo}, {Vigan}, {Claudi},
  {Alberi}, {Antichi}, {Baruffolo}, {Beuzit}, {Boccaletti}, {Bonnefoy},
  {Costille}, {Desidera}, {Dohlen}, {Fantinel}, {Feldt}, {Fusco}, {Giro},
  {Henning}, {Kasper}, {Langlois}, {Maire}, {Martinez}, {Moeller-Nilsson},
  {Mouillet}, {Moutou}, {Pavlov}, {Puget}, {Salasnich}, {Sauvage}, {Sissa},
  {Turatto}, {Udry}, {Vakili}, {Waters}, \& {Wildi}}]{2015A&A...576A.121M}
{Mesa}, D., {Gratton}, R., {Zurlo}, A., {et~al.} 2015, \aap, 576, A121

\bibitem[{{Milli} {et~al.}(2012){Milli}, {Mouillet}, {Lagrange}, {Boccaletti},
  {Mawet}, {Chauvin}, \& {Bonnefoy}}]{Milli2012}
{Milli}, J., {Mouillet}, D., {Lagrange}, A.~M., {et~al.} 2012, \aap, 545, A111

\bibitem[{{Miret-Roig} {et~al.}(2020){Miret-Roig}, {Galli}, {Brandner}, {Bouy},
  {Barrado}, {Olivares}, {Antoja}, {Romero-G{\'o}mez}, {Figueras}, \&
  {Lillo-Box}}]{2020A&A...642A.179M}
{Miret-Roig}, N., {Galli}, P.~A.~B., {Brandner}, W., {et~al.} 2020, \aap, 642,
  A179

\bibitem[{{Mo{\'o}r} {et~al.}(2017){Mo{\'o}r}, {Cur{\'e}}, {K{\'o}sp{\'a}l},
  {{\'A}brah{\'a}m}, {Csengeri}, {Eiroa}, {Gunawan}, {Henning}, {Hughes},
  {Juh{\'a}sz}, {Pawellek}, \& {Wyatt}}]{2017ApJ...849..123M}
{Mo{\'o}r}, A., {Cur{\'e}}, M., {K{\'o}sp{\'a}l}, {\'A}., {et~al.} 2017, \apj,
  849, 123

\bibitem[{{Mustill} \& {Wyatt}(2012{\natexlab{a}})}]{Mustill2012}
{Mustill}, A.~J. \& {Wyatt}, M.~C. 2012{\natexlab{a}}, \mnras, 419, 3074

\bibitem[{{Mustill} \& {Wyatt}(2012{\natexlab{b}})}]{2012MNRAS.419.3074M}
{Mustill}, A.~J. \& {Wyatt}, M.~C. 2012{\natexlab{b}}, \mnras, 419, 3074

\bibitem[{Nelder \& Mead(1965)}]{NelderMeade1965}
Nelder, J.~A. \& Mead, R. 1965, The Computer Journal, 7, 308

\bibitem[{{Padgett} \& {Stapelfeldt}(2016)}]{2016IAUS..314..175P}
{Padgett}, D. \& {Stapelfeldt}, K. 2016, in IAU Symposium, Vol. 314, IAU
  Symposium, ed. J.~H. {Kastner}, B.~{Stelzer}, \& S.~A. {Metchev}, 175--178

\bibitem[{{Parker} \& {Kavelaars}(2010)}]{2010ApJ...722L.204P}
{Parker}, A.~H. \& {Kavelaars}, J.~J. 2010, \apjl, 722, L204

\bibitem[{{Pawellek} {et~al.}(2014){Pawellek}, {Krivov}, {Marshall},
  {Montesinos}, {{\'A}brah{\'a}m}, {Mo{\'o}r}, {Bryden}, \&
  {Eiroa}}]{2014ApJ...792...65P}
{Pawellek}, N., {Krivov}, A.~V., {Marshall}, J.~P., {et~al.} 2014, \apj, 792,
  65

\bibitem[{{Pearce} {et~al.}(2021){Pearce}, {Beust}, {Faramaz}, {Booth},
  {Krivov}, {L{\"o}hne}, \& {Poblete}}]{2021arXiv210304977P}
{Pearce}, T.~D., {Beust}, H., {Faramaz}, V., {et~al.} 2021, arXiv e-prints,
  arXiv:2103.04977

\bibitem[{{Perrot} {et~al.}(2019){Perrot}, {Thebault}, {Lagrange},
  {Boccaletti}, {Vigan}, {Desidera}, {Augereau}, {Bonnefoy}, {Choquet}, {Kral},
  {Loh}, {Maire}, {M{\'e}nard}, {Messina}, {Olofsson}, {Gratton}, {Biller},
  {Brandner}, {Buenzli}, {Chauvin}, {Cheetham}, {Daemgen}, {Delorme}, {Feldt},
  {Lagadec}, {Langlois}, {Lannier}, {Mesa}, {Mouillet}, {Peretti},
  {Janin-Potiron}, {Salter}, {Sissa}, {Roux}, {Llored}, {Buey}, {Pavlov},
  {Weber}, \& {Petit}}]{2019A&A...626A..95P}
{Perrot}, C., {Thebault}, P., {Lagrange}, A.-M., {et~al.} 2019, \aap, 626, A95

\bibitem[{{Pinte} {et~al.}(2009){Pinte}, {Harries}, {Min}, {Watson},
  {Dullemond}, {Woitke}, {M{\'e}nard}, \&
  {Dur{\'a}n-Rojas}}]{2009A&A...498..967P}
{Pinte}, C., {Harries}, T.~J., {Min}, M., {et~al.} 2009, \aap, 498, 967

\bibitem[{{Pinte} {et~al.}(2006){Pinte}, {M{\'e}nard}, {Duch{\^e}ne}, \&
  {Bastien}}]{Pinte2006}
{Pinte}, C., {M{\'e}nard}, F., {Duch{\^e}ne}, G., \& {Bastien}, P. 2006, \aap,
  459, 797

\bibitem[{{Rameau} {et~al.}(2013){Rameau}, {Chauvin}, {Lagrange}, {Meshkat},
  {Boccaletti}, {Quanz}, {Currie}, {Mawet}, {Girard}, {Bonnefoy}, \&
  {Kenworthy}}]{2013ApJ...779L..26R}
{Rameau}, J., {Chauvin}, G., {Lagrange}, A.-M., {et~al.} 2013, \apjl, 779, L26

\bibitem[{{Rouleau} \& {Martin}(1991)}]{Rouleau1991}
{Rouleau}, F. \& {Martin}, P.~G. 1991, \apj, 377, 526

\bibitem[{{Soummer} {et~al.}(2012){Soummer}, {Pueyo}, \&
  {Larkin}}]{2012ApJ...755L..28S}
{Soummer}, R., {Pueyo}, L., \& {Larkin}, J. 2012, \apjl, 755, L28

\bibitem[{{Su} {et~al.}(2017){Su}, {MacGregor}, {Booth}, {Wilner}, {Flaherty},
  {Hughes}, {Phillips}, {Malhotra}, {Hales}, {Morrison}, {Ertel}, {Matthews},
  {Dent}, \& {Casassus}}]{2017AJ....154..225S}
{Su}, K. Y.~L., {MacGregor}, M.~A., {Booth}, M., {et~al.} 2017, \aj, 154, 225

\bibitem[{{Th{\'e}bault}(2009)}]{Thebault2009}
{Th{\'e}bault}, P. 2009, \aap, 505, 1269

\bibitem[{{Vigan} {et~al.}(2010){Vigan}, {Moutou}, {Langlois}, {Allard},
  {Boccaletti}, {Carbillet}, {Mouillet}, \& {Smith}}]{2010MNRAS.407...71V}
{Vigan}, A., {Moutou}, C., {Langlois}, M., {et~al.} 2010, \mnras, 407, 71

\bibitem[{{Wahhaj} {et~al.}(2016){Wahhaj}, {Milli}, {Kennedy}, {Ertel},
  {Matra}, {Boccaletti}, {del Burgo}, {Wyatt}, {Pinte}, {Lagrange}, {Absil},
  {Choquet}, {Gomez Gonzalez}, {Kobayashii}, {Mawet}, {Mouillet}, {Pueyo},
  {Dent}, {Augereau}, \& {Girard}}]{Wahhaj2016}
{Wahhaj}, Z., {Milli}, J., {Kennedy}, G., {et~al.} 2016, \aap, 596, L4

\bibitem[{{Weinberger} {et~al.}(2013){Weinberger}, {Anglada-Escud{\'e}}, \&
  {Boss}}]{2013ApJ...762..118W}
{Weinberger}, A.~J., {Anglada-Escud{\'e}}, G., \& {Boss}, A.~P. 2013, \apj,
  762, 118

\bibitem[{{Wisdom}(1980)}]{Wisdom1980}
{Wisdom}, J. 1980, \aj, 85, 1122

\bibitem[{{Wyatt}(2008)}]{Wyatt2008}
{Wyatt}, M.~C. 2008, \araa, 46, 339

\end{thebibliography}

\begin{appendix}
\section{Disks in Sco-Cen}
\label{Appendix:B}
We report in this section a homogeneous set of properties of debris disks that have been resolved in scattered light in Sco-Cen. The host star luminosities were estimated with the VOSA tool \citep{2008A&A...492..277B}. The dust radial distributions and the infrared luminosity are updated based on the Gaia-EDR3 parallaxes.

The parameters defining the disk radial distributions ($R_{in}$, $R_{out}$, $r_{0}$) vary depending on the framework used to model the images  \citep[e.g., ][]{Wahhaj2016, Esposito2020, 2016A&A...586L...8L}, or are directly extracted from the flux distribution in the images \citep[$r_{peak}$;][]{refId0} numerically deprojected whenever the disks are not seen edge-on.  We re-estimated  the minimum and maximum extent of the belts ($R_{in}$, $R_{out}$) from $r_{0}$, $\alpha_{in}$, and $\alpha_{out}$ whenever the former was not given in the reference papers following the method described in \cite{2021arXiv210706316A}. 





In the sample of targets, HD 98363 has been reported to form  a $\sim$7000 au wide  system with the K-type star Wray 15-788, bearing a circumstellar disk \citep{2019A&A...624A..87B}, making that system an exceptional case within the sample of disks that are resolved in Sco-Cen. However, the Gaia-EDR3 increases the distance difference between the two objects ($1.99\pm0.44$ pc) with respect to the DR2 ($1.08\pm0.84$pc), as well as the significance of the deviation in proper motion on the declination ($\mathrm{\Delta\:pm_{DEC}=0.598\pm0.025 \:mas/yr}$ for the EDR3 versus $\mathrm{\Delta\:pm_{DEC}=0.616\pm0.064\:mas/yr}$ for the DR2). Conversely, the  difference in proper motion in right ascension ($\mathrm{\Delta\:pm_{RA}=0.050\pm0.025\:mas/yr}$ for the EDR3 versus $\mathrm{\Delta\:pm_{RA}=0.092\pm0.068\:mas/yr}$ for the DR2) is slightly reduced. The nature of this putative system should be reinvestigated based on the improved astrometry and radial velocity measurement of both objects, which are provided as part of the future releases of Gaia.   

\newpage
 \begin{landscape}
 \begin{table}
\caption{Properties of debris disks in Sco-Cen inferred from scattered-light images and ranked according to the host star luminosity.}
\label{tab:dscocen}
\begin{center}
\footnotesize
\begin{tabular}{llllllllllllll}
\hline\hline
\#	Name	&		Distance	&  L/L$_{\odot}$ &  Sub-grp & \# belts &  $\mathrm{R_{in}}$ &   $\mathrm{R_{out}}$  & $\alpha_{in}$ & $\alpha_{out}$ & r$_{0}$ & $\mathrm{r_{peak}}$  & $\mathrm{L_{IR}/L_{\star}}\:^{c}$ & Morphology & Refs \\
		&		(pc)	 & & & &  (au) & (au) 	& & & (au)	&		(au) & & & \\
		
\hline
HD 156623$^{a}$ & $108.33\pm0.33$ & $11.73\pm0.50$ & UCL &   1  & $\leq$13.70 & 89.13 & \dots  & \dots & \dots & \dots  & $4.82\times 10^{-3}$ & ring & 2, 3  \\
HD 98363    &   $137.21\pm0.35$ & $9.95\pm0.10$ & LCC  & 1 & \dots  & \dots & \dots  & \dots & \dots  & \dots  & $9.71\times 10^{-4}$ & asym. ring & 4, 3  \\
HD 131835$^{a}$ & $129.74\pm0.47$ & $9.37\pm0.71$ & UCL   &   1 & 59.68 &	75.24  & \dots & \dots & \dots & 70.06  &   $3.43\times 10^{-3}$ &  ring & 5, 3\\
 &  &   &    &   2 	& 89.52	& 127.14 & $8.2^{+1.1}_{-0.8}$ & $-2.3\pm0.2$& $95.24^{+1.3}_{-1.2}$ & 103.79  &   $3.43\times 10^{-3}$ & ring  & 5, 3\\
HD 110058$^{a}$ & $130.08\pm0.53$ & $8.20\pm0.10$ & LCC & 1 & $\leq26$ & $\geq78$ & \dots& \dots& \dots & 39.02  & $2.98 \times 10^{-3}$ & asym. edge-on & 6, 3 \\
HD 143675 & $135.70\pm0.52$ & $7.98\pm0.43$ & UCL & 1 & $42.9^{+3.4}_{-7.4}$	& $50.8^{+1.4}_{-1.0}$ & \dots & \dots  & \dots & \dots  & $6.76\times10^{-4}$ & edge-on & 4, 3  \\
HD 111161 & $109.37\pm0.25$ & $7.81\pm0.08$ & LCC & 1 & $71.4^{0.5}_{-1.0}$ & $217.9^{+15.5}_{-15.3}$ & \dots & \dots  &  \dots & \dots  &  $5.06\times 10^{-3}$ & ring & 4, 3 \\
HD 106906$^{d}$ & $102.38\pm0.19$ & $5.92\pm0.36$ & LCC & 1   &   70.12$^{b}$ & 98.92$^{b}$ & 10.0 & $4.2\pm0.5$ & $77.19^{+3.28}_{-2.33}$ & \dots  & $5.01\times 10^{-3}$  & asym. ring & 7, 8, 3\\ 
HD 117214 & $107.35\pm0.25$ & $5.11\pm0.09$ & UCL & 1 & 43.48$^{b}$ & 56.04$^{b}$ & $24^{+18}_{-10}$ & $-4.2^{+0.3}_{-0.5}$ & $45.08\pm1.08$ & \dots  & $2.62\times10^{-3}$ & asym. ring & 2, 9, 3 \\
HD 115600 & $109.04\pm0.25$ & $4.57\pm0.07$ & LCC & 1  & 40.08$^{b}$ & 52.79$^{b}$ & 7.5 & -7.5 & $46\pm2$ & \dots  &   $2.60\times10^{-3}$ & ring &  10, 11, 3\\
HD 120326$^{d}$ & $113.27\pm0.38$ & $4.15\pm0.05$ & UCL & 1 & 55.99$^{b}$ & 75.66$^{b}$ & 10:  & -5: & $61.8\pm3.2$ & \dots  & $1.53\times 10^{-3}$ & ring & 12,  3 \\ 
        &       &       &       &   2   &  123.75$^{b}$ & 155.97$^{b}$ & 10: &  $-8^{+4}_{-8}$ & 137.1$\pm8.5$  & \dots  & $1.53\times 10^{-3}$ & asym. ring & 12, 3 \\ 
HD 114082$^{a}$ & $95.06\pm0.20$ & $3.50\pm0.04$  & LCC & 1  & 30.43$^{b}$ & 41.25$^{b}$ & 10: & -4.98 & 30.08  &  \dots  & $4.92\times 10^{-3}$ & ring &  13, 3 \\
HD 141011 & $128.38\pm0.32$ & $3.39\pm0.03$ & UCL & 1  &   121.17$^{b}$  &   137.31$^{b}$ & 20: & $-13.9^{+3.4}_{-4.6}$ & $127.5\pm3.9$ & \dots  & $2.90\times10^{-4}$ & ring & 6, 3 \\
HD 146897$^{a}$ & $132.19\pm0.42$ & $3.27\pm0.06$ & US & 1 & 64.40$^{b}$ & 117.52$^{b}$ & $5.0\pm2.8$ & $-2.5\pm1.4$ & $78\pm18$ & \dots  & $1.06\times10^{-2}$ & edge-on & 14, 3 \\
HD 145560$^{a}$ & $121.23\pm0.29$ & $3.18\pm0.14$ & UCL & 1 & $69.1^{+3.0}_{-1.3}$ & $225.7^{+27.4e}_{-10.9}$ & \dots & \dots & \dots & \dots  & $1.38\times 10^{-3}$ & ring & 1, 3, 4\\ 
HD 129590$^{a}$ & $136.32\pm0.44$ & $2.92\pm0.56$ & UCL & 1 & 42.18$^{b}$ & 122.30$^{b}$ & $3.15\pm0.03$ & $-1.31\pm0.01$ & $57.3\pm0.2$ & \dots  & $7.97\times10^{-3}$ & ring & 15, 3 \\
HD 111520$^{a, d}$ & $108.05\pm0.21$ & $2.51\pm0.17$ & LCC &  1 & $\leq$32 & $\geq108$  & \dots & \dots & \dots & 62   & $1.10\times10^{-3}$ & asym. edge-on & 16, 3 \\
\hline
\end{tabular}
\end{center}
\tablefoot{References: 1 - this work, 2 - \cite{Esposito2020}}, 3 - \cite{2020yCat.1350....0G}, 4 - \cite{Hom2020}, 5 - \cite{refId0}, 6 - \cite{2015ApJ...812L..33K}, 7 - \cite{2015ApJ...814...32K}, 8 - \cite{2016A&A...586L...8L}, 9 - \cite{2020A&A...635A..19E},  10 - \cite{Currie2015}, 11 - \cite{2019AJ....157...39G}, 12 - \cite{Bonnefoy2017}, 13 - \cite{Wahhaj2016}, 14 - \cite{2017A&A...607A..90E}, 15 - \cite{2017ApJ...843L..12M}, 16 - \cite{2016arXiv160502771D}. \\
Notes: $^{a}$ Also resolved at submillimeter wavelengths \citep{2016ApJ...828...25L, 2020MNRAS.497.2811K}. $^{b}$ Estimated from $R_{0}$, $\alpha_{in}$, and $\alpha_{out}$. $^{c}$ Fractional luminosity of both belts, $^{d}$ Extended structures at optical wavelengths \citep{2015ApJ...814...32K, 2016IAUS..314..175P, Bonnefoy2017}, $^{e}$ \cite{Esposito2020}  presents a lower limit for Rout of 196.2 au.\\
\end{table}
\end{landscape}

\section{Disk modeling}
\label{app_disk_modelling}

\begin{figure*}
\begin{center}
\includegraphics[width=\linewidth]{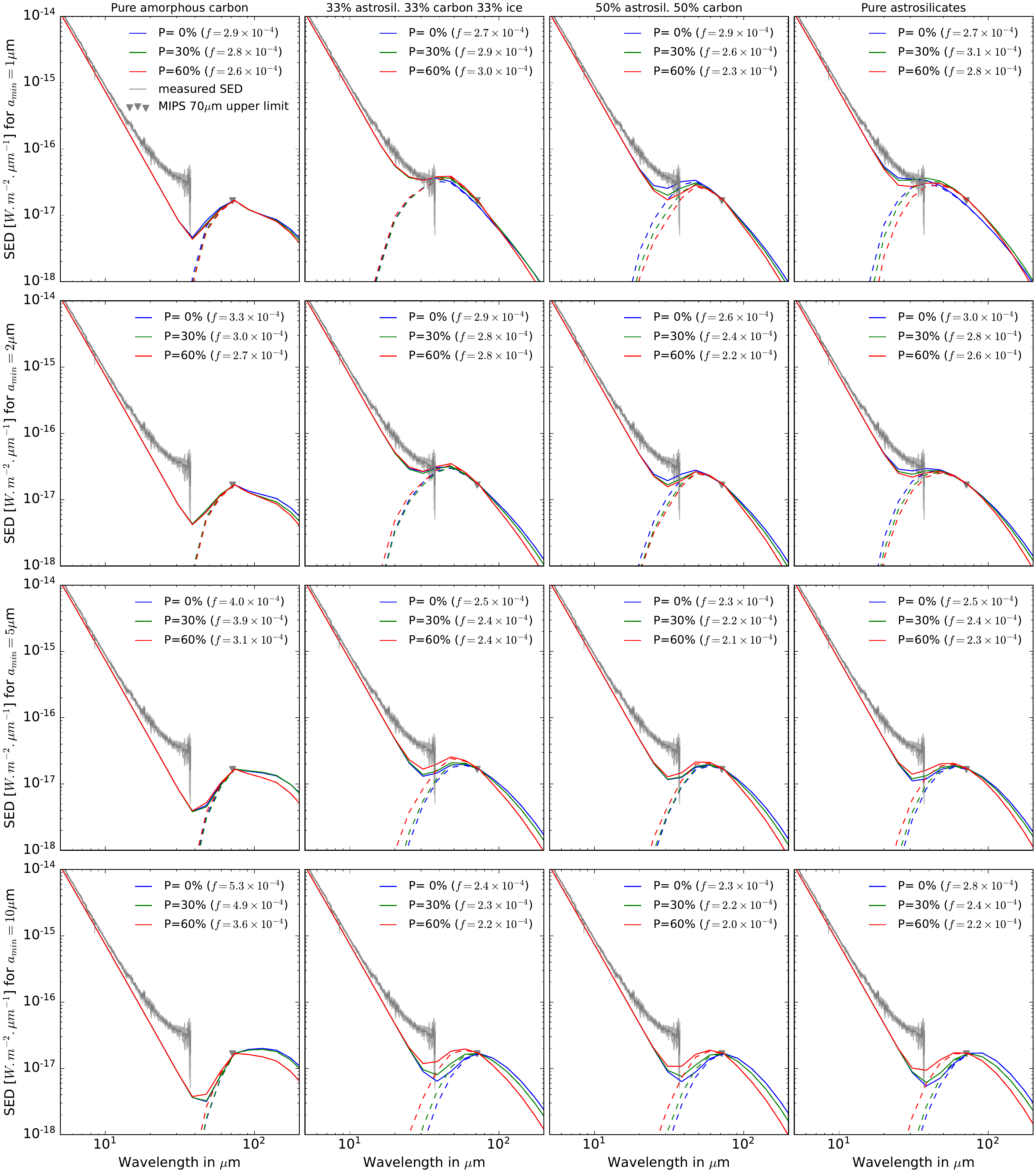}
\caption{Spectral energy distribution of the 48 disk models (four different compositions in columns, four different minimum particle sizes in rows, and three different porosities in three colors). The infrared excess of each model is indicated in the label.}
\label{fig_grid_sed}
\end{center}
\end{figure*}

Fig. \ref{fig_grid_sed} shows the SED of the 48 disk models.

\end{appendix}

\end{document}